\documentclass[prd,amsfonts,amssymb,noshowpacs,nofootinbib,preprintnumbers]{revtex4}
\pdfoutput=1 
\usepackage{amsmath,graphicx,hyperref,epsfig,soul,subfig,color,adjustbox}

\begin{document}

\title{Large-scale structure in mimetic Horndeski gravity}
\author{Frederico Arroja$^{1}$\footnote{arroja{}@{}phys.ntu.edu.tw}, Teppei Okumura$^{2}$\footnote{tokumura@asiaa.sinica.edu.tw}, Nicola Bartolo$^{3,4,5}$\footnote{nicola.bartolo@pd.infn.it}, Purnendu Karmakar$^{3,4}$\footnote{purnendu.karmakar@pd.infn.it} and Sabino Matarrese$^{3,4,5,6}$\footnote{sabino.matarrese@pd.infn.it}
}
\affiliation{
{}$^{1}$Leung Center for Cosmology and Particle Astrophysics, National Taiwan University, Taipei 10617, Taiwan
\\
{}$^{2}$Institute of Astronomy and Astrophysics, Academia Sinica,
P. O. Box 23-141, Taipei 10617, Taiwan
\\
{}$^{3}$Dipartimento di Fisica e Astronomia ``G. Galilei'', Universit\`{a} degli Studi di
Padova, via Marzolo 8,  I-35131 Padova, Italy
\\
{}$^{4}$INFN, Sezione di Padova, via Marzolo 8,  I-35131 Padova, Italy
\\
{}$^{5}$INAF - Osservatorio Astronomico di Padova, Vicolo dell'Osservatorio 5, I-35122 Padova, Italy
\\
{}$^{6}$Gran Sasso Science Institute, INFN, Viale F. Crispi 7,  I-67100 L'Aquila, Italy
}

\begin{abstract}
In this paper, we propose to use the mimetic Horndeski model as a model for the dark universe. Both cold dark matter (CDM) and dark energy (DE) phenomena are described by a single component, the mimetic field. In linear theory, we show that this component effectively behaves like a perfect fluid with zero sound speed and clusters on all scales. For the simpler mimetic cubic Horndeski model, if the background expansion history is chosen to be identical to a perfect fluid DE (PFDE) then the mimetic model predicts the same power spectrum of the Newtonian potential as the PFDE model with zero sound speed. In particular, if the background is chosen to be the same as that of LCDM, then also in this case the power spectrum of the Newtonian potential in the mimetic model becomes indistinguishable from the power spectrum in LCDM on linear scales. A different conclusion may be found in the case of non-adiabatic perturbations. We also discuss the distinguishability, using power spectrum measurements from LCDM N-body simulations as a proxy for future observations, between these mimetic models and other popular models of DE. For instance, we find that if the background has an equation of state equal to -0.95 then we will be able to distinguish the mimetic model from the PFDE model with unity sound speed. On the other hand, it will be hard to do this distinction with respect to the LCDM model.
\end{abstract}


\date{\today}
\maketitle

\section{Introduction\label{sec:intro}}

It is widely believed that in order to explain observations of the cosmic microwave background (CMB) and the large scale structure (LSS) of the universe (in addition to several other astrophysical observations), we need to introduce two unknown components into the energy density budget, if one assumes that the theory of gravity is described by Einstein's General Relativity (GR) \cite{Dodelson:2003ft,Weinberg:2008zzc}. These two components go by the names of dark matter (DM) and dark energy (DE).

One of the simplest cosmological models, the $\Lambda$-Cold-Dark-Matter model (LCDM), fits reasonably well all present observations with only six cosmological parameters and it has become the so-called standard model of cosmology \cite{Adam:2015rua,Ade:2015xua}. In this spatially flat model, the dark universe, i.e. the DM and the DE components, is described by a cold collisionless fluid, the CDM component, and a cosmological constant (CC), $\Lambda$, that drives the current observed accelerated expansion \cite{Riess:1998cb,Perlmutter:1998np}. The energy density of the dark universe amounts to about $95\%$ of the total energy density of the Universe today.
This simplest model is not without conceptual problems (e.g. see \cite{Weinberg:1988cp}). For example, the nature of the DM is still unknown and the cosmological constant, if it is interpreted as receiving a contribution from the vacuum energy of the elementary particles of the Standard Model (SM) of particle physics, has an observed magnitude that is many orders of magnitude below the expected value.

The difficulties related to the CC have motivated the community to look for alternative models to explain the acceleration of the universe \cite{Bull:2015stt}. For example, the DE could be described by a perfect fluid with an equation-of-state (EOS) and a sound speed or by some scalar field usually called quintessence \cite{Tsujikawa:2013fta}. An alternative class of explanations is to modify the theory of gravity with respect to GR (see e.g., \cite{Frieman:2008sn,Clifton:2011jh,Joyce:2016vqv} for reviews).
While in the LCDM model, the CDM is described by a perfect fluid with zero pressure, other alternative explanations have also been proposed, e.g. axion-like particles \cite{Hui:2016ltb} and mimetic DM \cite{Chamseddine:2013kea}. The latter will be the main focus of this work.

This new mimetic DM model \cite{Chamseddine:2013kea,Chamseddine:2014vna} (see also \cite{Lim:2010yk}), has attracted considerable attention because it was shown that it can mimic the behavior of CDM even in the absence of any other form of matter. It is a modification of GR in that the theory becomes a scalar-tensor theory. With a small generalization, the mimetic field can mimic the behavior of almost any type of matter and one can have almost any desired background expansion history \cite{Chamseddine:2014vna}. Many different aspects of these mimetic theories have been studied. See the review \cite{Sebastiani:2016ras} and references therein.

In \cite{Arroja:2015wpa}, we proposed the mimetic Horndeski gravity model. This is a fairly general scalar-tensor gravity theory which generalizes the original mimetic DM model and contains many of the different mimetic gravity models present in the literature.
This mimetic theory is fairly general because the starting point for its construction was Horndeski's gravity \cite{Horndeski:1974wa} which is a very general theory itself. However, recently, generalizations of Horndeski's gravity, which propagate the same number of degrees of freedom, have been found (see for example \cite{Langlois:2017mdk} and references therein).

In \cite{Arroja:2015yvd}, we performed further studies of scalar linear perturbations in the mimetic Horndeski gravity model and showed that their sound speed is exactly zero. This theory represents also a modification of gravity with respect to GR and it contains non-minimal couplings. However, we should say that it is not the most general mimetic theory possible. In fact, in the early work \cite{Chamseddine:2014vna}, a term proportional to $(\Box\varphi)^2$ was added to the action of the mimetic DM theory in order to have a non-zero sound speed for the scalar perturbations. The model that includes this term (and other models with higher-derivatives terms) is not included in the mimetic Horndeski model. They possess interesting features and have problems on their own. One of the problems is, for example, the presence of instabilities \cite{Ramazanov:2016xhp}. These instabilities are either ghost-like or gradient instabilities. Ref. \cite{Ramazanov:2016xhp} showed that the ghost instability can be made harmless by a suitable choice of the sound speed. Nevertheless, one of the interesting features of the model in \cite{Chamseddine:2014vna} is that it is equivalent \cite{Ramazanov:2016xhp} to the infrared limit of the projectable version of Ho\v{r}ava-Lifshitz gravity \cite{Horava:2009uw}, which is one of the candidates for the theory of quantum gravity. It was also shown that this model is a viable DM model if the sound speed is extremely small \cite{Capela:2014xta,Ramazanov:2016xhp} and in this sense we expect that the models studied in this work, which have exactly zero sound speed, and some of our conclusions will also be useful in terms of the phenomenology of that and related models.

In this work, we will use the mimetic Horndeski gravity as a theory of unified DM and DE (see Ref. \cite{Bertacca:2010ct} for a review), dubbed unified dark matter\footnote{Using a so-called $\lambda\varphi$-fluid, which turns out to be a mimetic model, Ref. \cite{Lim:2010yk} was the first to propose a model of unified dark matter within the context of mimetic theories.}. Similarly to those unified dark matter models, here, the mimetic field will describe the entire dark universe with just one entity. This represents a minimalistic approach in our view because we use only one component instead of two to describe the dark universe. And it is a reasonable approach given that there is no evidence to support that DM and DE (if not explained as a CC) are distinct entities. Our use of mimetic gravity assumes (although this is not a necessity) that the mimetic field is only gravitationally coupled to the fields of the SM. While this is a standard implicit assumption for DE, it is not the usual assumption made for DM. In fact, currently there are many experiments on Earth searching for the first direct DM detection. If the true model of DM is mimetic DM, or many of the axion-like models in the literature, then all these direct detection experiments are bound to fail. This said, it is important to stress that mimetic DM might be only one of the many components of DM and the other components could be detectable on Earth. We also note that mimetic gravity theory, like other scalar-tensor theories, may have many uses and in particular may be used to model DE only, or DM only or other phenomena like for instance inflation in the very early Universe.

In this paper, we compute consequences of these models for the LSS of the universe. We will show that they may, depending on model parameters, give reasonable linear structure formation predictions which are in good agreement with some current observations and will certainly be stringently tested with future LSS surveys, such as the Subaru Prime Focus Spectrograph (PFS) \cite{Ellis:2012rn}, the Dark Energy Spectroscopic
Instrument (DESI) \cite{Aghamousa:2016zmz}, and the EUCLID mission \cite{Amendola:2016saw}\footnote{http://sci.esa.int/euclid/}. The linear growth rate of the matter density contrast in a mimetic model of unified DM and DE has been computed before in \cite{Matsumoto:2016rsa}. The model used there is a particular case of the more general models discussed in the present paper.

This paper is organized as follows. In the next section, we briefly introduce the mimetic Horndeski model and our notation which is mostly the same as in \cite{Arroja:2015yvd}. In Section \ref{sec:equiv}, we discuss the relation of our models to perfect fluid models for the dark universe.
In Section \ref{sec:numerical}, we describe our numerical code to integrate the evolution equations and the power spectrum measurements from N-body LCDM simulations that we will use. We also show a comparison between the accuracy of our simple code with both well-known fitting functions and the full Boltzmann-Einstein system numerical integrator, CAMB \cite{Lewis:1999bs}\footnote{http://camb.info/}.
In Section \ref{sec:distinguish}, using the results from the N-body simulations as a proxy for future observations, we shall discuss the distinguishability between our simple mimetic models and popular DE models like the LCDM model and the perfect fluid dark energy (PFDE) model.
Section \ref{sec:con} is devoted to the conclusions.
In three appendices, we present a second-order evolution equation for the Newtonian potential, $\Phi$, in mimetic gravity coupled with a fluid (Appendix \ref{app:evoleq}), we review the equations of motion in PFDE models that are to be integrated numerically (Appendix \ref{app:eomPFDE}) and finally we review the computation of the power spectrum of $\Phi$ in clustering PFDE models that is to be used in several plots in the main text (Appendix \ref{app:PSPFDE}). In this paper we will use the mostly plus metric signature. Greek indices denote spacetime coordinate labels and run from 0 to 3, with 0 denoting the time coordinate. Latin indices denote three-space coordinates and run from 1 to 3. The reduced Planck mass is $M_{Pl}=1/\sqrt{8\pi G}$, where $G$ is Newton's constant.

\section{The mimetic Horndeski gravity and notation\label{sec:MHN}}

In this section, we briefly introduce the mimetic Horndeski gravity model that was first proposed in \cite{Arroja:2015wpa}. In this work, we shall use this model as a unified dark matter model. Then we shall discuss linear scalar perturbations, first studied in \cite{Arroja:2015yvd}. We will mostly follow the notation of \cite{Arroja:2015yvd}.

For a very general action of mimetic gravity,
\begin{eqnarray}
S&=&\int d^4x\sqrt{-g}\mathcal{L}[g_{\mu\nu},\partial_{\lambda_1}g_{\mu\nu},\ldots,\partial_{\lambda_1}\ldots\partial_{\lambda_p}g_{\mu\nu},\varphi,\partial_{\lambda_1}\varphi,\ldots,\partial_{\lambda_1}\ldots\partial_{\lambda_q}\varphi]+S_m[g_{\mu\nu},\phi_m]\nonumber\\
&+&\int d^4x\sqrt{-g}\lambda\left(b(\varphi)g^{\mu\nu}\partial_\mu\varphi\partial_\nu\varphi-1\right)
,\label{generalaction}
\end{eqnarray}
where $\varphi$ is the mimetic scalar field, $\lambda$ is a Lagrange multiplier field, $\phi_m$ is a generic matter field with action $S_m$ which is coupled with the metric $g_{\mu\nu}$ only. $b(\varphi)$ is a potential function and the integers $p,q\geq2$.
By defining the following quantities
\begin{eqnarray}
E^{\mu\nu}=\frac{2}{\sqrt{-g}}\frac{\delta(\sqrt{-g}\mathcal{L})}{\delta g_{\mu\nu}},
T^{\mu\nu}=\frac{2}{\sqrt{-g}}\frac{\delta (\sqrt{-g}\mathcal{L}_m)}{\delta g_{\mu\nu}}, \,
\Xi_m=\frac{\delta (\sqrt{-g}\mathcal{L}_m)}{\delta \phi_m},
\,\mathrm{where}\,\, S_m[g_{\mu\nu},\phi_m]=\int d^4x\sqrt{-g}\mathcal{L}_m[g_{\mu\nu},\phi_m],
\end{eqnarray}
where $\mathcal{L}_m$ is the matter Lagrangian density and $T^{\mu\nu}$ denotes its energy-momentum tensor,
one can write a complete set of equations of motion as \cite{Arroja:2015wpa,Arroja:2015yvd}
\begin{eqnarray}
&&
b(\varphi)g^{\mu\nu}\partial_\mu\varphi\partial_\nu\varphi-1=0,\label{normalization2}\\
&&
E^{\mu i}+T^{\mu i}=(E+T)b(\varphi)\partial^\mu\varphi\partial^i\varphi,\label{mEE2}\\
&&
\Xi_m=0\label{mMF}.
\end{eqnarray}
The first equation is known as the mimetic constraint and one sees that the time-time metric equation of motion is redundant with respect to the previous set \cite{Arroja:2015yvd}. Furthermore, the mimetic scalar field equation of motion is also redundant \cite{Arroja:2015yvd}. Eq. (\ref{mMF}) implies the conservation of the energy momentum tensor $\nabla_\mu T^{\mu\nu}=0$. The Lagrange multiplier field is given by $\lambda=(E+T)/2$, where $E$ and $T$ are the traces of $E^{\mu\nu}$ and $T^{\mu\nu}$ respectively.
The Lagrangian $\mathcal{L}$ is the Horndeski Lagrangian \cite{Horndeski:1974wa,Deffayet:2011gz,Kobayashi:2011nu} which is given by the sum of the following four terms
\begin{eqnarray}
\mathcal{L}_0 & = & K\left(X,\varphi\right),\\
\mathcal{L}_1 & = & -G_3\left(X,\varphi\right)\Box\varphi,\\
\mathcal{L}_2 & = & G_{4,X}\left(X,\varphi\right)\left[\left(\Box\varphi\right)^{2}-\left(\nabla_{\mu}\nabla_{\nu}\varphi\right)^{2}\right]
+R\,G_4\left(X,\varphi\right) ,\\
\mathcal{L}_3 & = & -\frac{1}{6}G_{5,X}\left(X,\varphi\right)\left[\left(\Box\varphi\right)^{3}-3\Box\varphi\left(\nabla_{\mu}\nabla_{\nu}\varphi\right)^{2}
+2\left(\nabla_{\mu}\nabla_{\nu}\varphi\right)^{3}\right]+G_{\mu\nu}\nabla^{\mu}\nabla^{\nu}\varphi\,
G_5\left(X,\varphi\right),
\end{eqnarray}
where $X=-1/2\nabla_\mu\varphi\nabla^\mu\varphi$, $(\nabla_\mu\nabla_\nu\varphi)^2=\nabla_\mu\nabla_\nu\varphi\nabla^\mu\nabla^\nu\varphi$ and $(\nabla_\mu\nabla_\nu\varphi)^3=\nabla_\mu\nabla_\nu\varphi\nabla^\mu\nabla^\rho\varphi\nabla^\nu\nabla_\rho\varphi$.
The subscripts $,\varphi$ and $,X$ denote derivatives with respect to $\varphi$ and $X$ respectively. The Horndeski functions $K,\,G_3,\,G_4,\,G_5$ of the two variables, $X$ and $\varphi$, define a particular (mimetic) Horndeski theory. For a general mimetic Horndeski model, the function $b(\varphi)$ can be reabsorbed by a field redefinition of $\varphi$ without losing generality \cite{Arroja:2015wpa}. Because the Horndeski theory is form invariant under the field redefinition, it just amounts to consider a different original starting set of functions $K,\,G_3,\,G_4,\,G_5$. Furthermore, this field redefinition would not change the physical consequences of the theory. In this work, we decided to keep the general function $b(\varphi)$ in the equations for the sake of comparison with previous works \cite{Lim:2010yk,Deruelle:2014zza,Arroja:2015wpa,Arroja:2015yvd}.

We now turn to the study of linear scalar perturbations in this model with the matter field being described by a fluid. All the necessary background and perturbed equations of motion can be found in Appendices A, B and C of \cite{Arroja:2015yvd}. Here we will briefly present only the equations which we will need to use later.

We will work in the Poisson gauge, neglect vector and tensor perturbations and assume a spatially flat Friedmann-Lema\^{i}tre-Robertson-Walker (FLRW) background. The metric is then written as
\begin{equation}
g_{00}=-a^2(\tau)\left(1+2\Phi\right),\quad g_{0i}=0, \quad g_{ij}=a^2(\tau)\left(1-2\Psi\right)\delta_{ij},
\end{equation}
where $a$ is the FLRW scale factor that depends on the conformal time $\tau$, $\Phi$ denotes the generalised Newtonian (Bardeen) potential and $\Psi$ the curvature perturbation. The scalar field is expanded as $\varphi(\tau,\mathbf{x})=\bar{\varphi}(\tau)+\delta\varphi(\tau,\mathbf{x})$, where $\bar{\varphi}$ denotes the background field value and $\delta\varphi$ is its perturbation. A prime denotes derivative with respect to conformal time. In this work, a bar over a quantity denotes the background value. We chose this notation, different from that in \cite{Arroja:2015yvd}, because here we want to use the subscript 0 to denote a quantity at the present day.

The fluid has an energy-momentum tensor of the form
\begin{equation}
T_{\mu\nu}=(\rho+p)u_\mu u_\nu+pg_{\mu\nu}+\pi_{\mu\nu},
\end{equation}
where $\rho$ is the energy density, $p$ the pressure and $\pi_{\mu\nu}$ is the anisotropic stress tensor which vanishes for perfect fluids. The four-velocity, $u^\mu$, is a time-like vector and obeys the constraint $u_\mu u^\mu=-1$. It can be used to find
\begin{equation}
u^0=a^{-1}(1-\Phi),\quad u^i=a^{-1}v^i,
\end{equation}
where the velocity $v^i$, is a first-order quantity, and can be written in terms of a scalar quantity, $v$, as $v^i=\delta^{ij}\partial_j v$ (because we neglect intrinsic vector perturbations). Similarly, the anisotropic stress tensor can be described by a scalar, denoted by $\Pi$ (see \cite{Arroja:2015yvd} for all the details).
Then the background equations are simply
\begin{equation}
-a^{-2}b(\bar{\varphi})(\bar{\varphi}')^2=1,\qquad \bar{E}_{ij}=-a^2 \bar{p}\delta_{ij}, \qquad \bar{\rho}'+3\mathcal{H}(\bar{\rho}+\bar{p})=0,\label{eq0}
\end{equation}
with the definition $\mathcal{H}\equiv a'/a$.
At first order, the set of independent equations is
\begin{eqnarray}
&&2\bar{b}\delta\varphi'+\bar{\varphi}'\bar{b}_{,\varphi}\delta\varphi-2\bar{b}\bar{\varphi}'\Phi=0,\label{eq1}
\\
&&f_7\Psi+f_8\delta\varphi+f_9\Phi+a^2\Pi=0,\label{eq2}
\\
&&f_{10}\Psi'+f_{11}\delta\varphi'+\left(f_{20}+\frac{a^2(\bar{E}+\bar{T})}{\bar{\varphi}'}\right)\delta\varphi+f_{14}\Phi-a^2\left(\bar{\rho}+\bar{p}\right)v=0,\label{eq4}
\\
\nonumber
\\
&&\delta\rho'+3\mathcal{H}(\delta\rho+\delta p)-3(\bar{\rho}+\bar{p})\Psi'+(\bar{\rho}+\bar{p})\partial^2v=0,\label{eq5}
\\
&&
\left((\bar{\rho}+\bar{p})v\right)'+\delta p+\frac{2}{3}\partial^2\Pi+4\mathcal{H}(\bar{\rho}+\bar{p})v+(\bar{\rho}+\bar{p})\Phi=0,\label{eq6}
\end{eqnarray}
where $\bar{E}_{ij}$ and the $f_i$ functions are complicated functions of $K$, $G_3$, $G_4$, $G_5$ and their derivatives. The long expressions for $\bar{E}_{ij}$ and $f_i$ together with useful identities that these functions obey can be found in Appendices A and B of \cite{Arroja:2015yvd}. In Appendix \ref{app:evoleq}, using the previous equations, we derive a second order evolution equation for $\Phi$. This generalizes results present in \cite{Arroja:2015yvd} to include the coupling to a fluid.
The last two equations in the set are a direct consequence of the conservation of the energy-momentum tensor and thus are valid independently of the theory of gravity. For the case of multiple non-interacting fluids, these equations are valid with the definitions $\bar{\rho}\equiv\sum_\mathfrak{f}\bar{\rho}_\mathfrak{f}$, $\delta\rho\equiv\sum_\mathfrak{f}\delta\rho_\mathfrak{f}$, where the subscript $\mathfrak{f}$ denotes the different fluids, and equally for the pressure. $(\bar{\rho}+\bar{p})v\equiv\sum_\mathfrak{f}(\bar{\rho}_\mathfrak{f}+\bar{p}_\mathfrak{f})v_\mathfrak{f}$. Furthermore, in this case, these last two equations are also valid for each individual fluid.
In the standard model of cosmology there are several particles that are important for the evolution of the universe and here we will follow a hydrodynamical approach, namely approximate them as perfect fluids. The label that specifies the fluid type, $\mathfrak{f}$, takes the following values: $\gamma$ for photons, $\nu$ for massless neutrinos (and anti-neutrinos), $r$ for radiation (sum of the photons and the neutrinos), $b$ for baryons, $CDM$ for cold dark matter, $m$ for matter (sum of the baryon and the CDM) and PFDE for perfect fluid dark energy. Perfect fluids have $\Pi_\mathfrak{f}=0$ (note however that these assumptions are not always accurate for photons and free-streaming neutrinos). Radiation has $w_r=1/3$, matter has $w_m=0$ and in our numerical code we allow the equation of state (EOS) of the DE to be time dependent following the so-called Chevallier-Polarski-Linder (CPL) parametrization as $w(a)=w_0+w_a(1-a)$ \cite{Chevallier:2000qy,Linder:2002et}.

The EOS for the fluid $\mathfrak{f}$ is $w_\mathfrak{f}\equiv\bar{p}/\bar{\rho}$. At the linear level, the pressure can be written as (see e.g. \cite{Kodama:1985bj,Bean:2003fb})
\begin{equation}
\delta p_\mathfrak{f}=c_{(\mathfrak{f})a}^2\delta\rho_\mathfrak{f}+\left(c_{(\mathfrak{f})s}^2-c_{(\mathfrak{f})a}^2\right)\left(\delta\rho_\mathfrak{f}+\bar{\rho}'_\mathfrak{f} v_\mathfrak{f}\right),
\end{equation}
where the adiabatic sound speed is defined as $c_{(\mathfrak{f})a}^2\equiv\bar{p}_\mathfrak{f}'/\bar{\rho}_\mathfrak{f}'$ and it is $c_{(\mathfrak{f})a}^2=w_\mathfrak{f}$ if $w_\mathfrak{f}$ is a constant, and the sound speed is denoted as $c_{(\mathfrak{f})s}$, which is a new parameter independent of $c_{(\mathfrak{f})a}$. For all these fluids we take $c_{(\mathfrak{f})s}=c_{(\mathfrak{f})a}$ except for the DE fluid.

We define the density contrast as $\delta_\mathfrak{f}\equiv\delta\rho_\mathfrak{f}/\bar{\rho}_\mathfrak{f}.$
The energy density parameter is defined as usual
\begin{equation}
\Omega_\mathfrak{f}=\frac{\bar{\rho}_\mathfrak{f}}{3H^2M_{Pl}^2},
\end{equation}
where $H\equiv \dot a/a$, the dot denotes derivative with respect to cosmic time t. We define a rescaled velocity as $\tilde{v}_\mathfrak{f}=\mathcal{H}v_\mathfrak{f}$.

We define the Fourier transform of some quantity $Q(\tau,\mathbf{x})$ as $Q(\tau,\mathbf{x})=1/(2\pi)^3\int d^3kQ(\tau,\mathbf{k})e^{i\mathbf{k}\cdot\mathbf{x}}$, the inverse is $Q(\tau,\mathbf{k})=\int d^3xQ(\tau,\mathbf{x})e^{-i\mathbf{k}\cdot\mathbf{x}}$. In canonical single-field inflation, the comoving gauge (also called uniform-field gauge or unitary gauge) is defined by setting the inflaton perturbation to zero and the comoving curvature perturbation is then the perturbation to the three-dimensional spatial metric as $h_{ij}=a^2\exp\left(2\mathcal{R}\right)\delta_{ij}$.
The power spectrum of $\Phi$ is defined as
\begin{equation}
P_\Phi(\tau,k)=\frac{2 k^3}{(2\pi)^2}|\Phi(\tau,k)|^2.
\end{equation}
In this work we are mainly interested in computing the predictions for the power spectrum of $\Phi$ in mimetic models. $\Phi$ is well-known to be gauge invariant and can be directly related to gravitational lensing observables for example. It can also be related with observables in LSS surveys, like for instance the power spectrum of galaxies, by using a suitable bias model. A systematic derivation of bias in clustering dark energy and mimetic models is still an open question and is currently under investigation.

The primordial power spectrum of the comoving curvature perturbation, $\Delta_\mathcal{R}^2(k)$, is
\begin{equation}
\Delta_\mathcal{R}^2(k)=\frac{2 k^3}{(2\pi)^2}|\mathcal{R}(\tau_k,k)|^2=\Delta_\mathcal{R}^2(k_0)\left(\frac{k}{k_0}\right)^{n_s-1}, \label{primordialspectrum}
\end{equation}
where $\tau_k$ is the horizon crossing time (after horizon exit $\mathcal{R}$ becomes constant in single-field slow-roll inflation) for the wavemode $k$ and it is related with the two-point correlation function as
\begin{equation}
\langle\mathcal{\hat{R}}^2(\mathbf{x})\rangle=\int \frac{dk}{k}\Delta_\mathcal{R}^2(k).
\end{equation}

\section{Equivalent predictions\label{sec:equiv}}

In this section, we will compare the mimetic Horndeski gravity, as a model of unified dark matter, with perfect fluid models for the dark universe assuming that gravity is described by GR. The equations of motion for these latter models can be found in Appendix \ref{app:eomPFDE}.

As we have seen in the previous section, the relevant equations in the mimetic Horndeski model are Eqs. (\ref{eq1})-(\ref{eq6}).
It turns out that by defining a scalar field velocity
\begin{equation}
v_\varphi\equiv-\frac{\delta\varphi}{\bar{\varphi}'},
\end{equation}
one can write Eq. (\ref{eq1}) in a very suggestive way as
\begin{equation}
v_\varphi'+\mathcal{H}v_\varphi+\Phi=0.\label{dustv}
\end{equation}
Comparing this equation with Eq. (\ref{vp}) one can see that they are equal if $c_{(\mathfrak{f})s}=\Pi_\mathfrak{f}=0$, that is, the previous equation is the equation for the velocity of dust (recall that we neglect vector and tensor perturbations).

After a few manipulations, in particular using the mimetic constraint, the background equations and the previously mentioned identities for the $f_i$ functions, Eq. (\ref{eq4}) can be written as
\begin{equation}
\Psi'+\mathcal{H}\Phi+\left(\frac{a^2\left(\bar{\rho}+\bar{p}\right)}{f_{10}}+\mathcal{H}^2-\mathcal{H}'\right)v_\varphi-\frac{a^2}{f_{10}}\left(\bar{\rho}+\bar{p}\right)v=0.
\label{mc}
\end{equation}
The previous three equations together with Eq. (\ref{eq2}) form a closed system. Since we are interested in introducing some additional matter fields, in particular radiation and baryons, we need to add to the system of equations their equations of motion given by Eqs. (\ref{deltap}) and (\ref{vp}). It is important to note that we want to use these mimetic models as unified dark matter models so in this section $\mathfrak{f}$ denotes radiation (photons plus massless neutrinos) and baryons only. We do not introduce CDM or DE by hand. The previous three equations are simple and can be used to solve for $\Phi$, $\Psi$ and $v_\mathfrak{f}$ for a general mimetic Horndeski model once we specify the Horndeski functions.

In the rest of the paper, for simplicity, we will consider less general models as
\begin{equation}
G_4=\frac{M_{Pl}^2}{2}, \quad G_5=0,
\label{models}
\end{equation}
while the functions $b$, $K$ and $G_3$ are still kept general. This choice switches off the non-minimal coupling in the Horndeski Lagrangian.
We call these models mimetic cubic Horndeski and in the literature they are also often known as ``kinetic gravity braiding" \cite{Deffayet:2010qz}. They are still very general, and include the original models in \cite{Chamseddine:2013kea,Chamseddine:2014vna,Lim:2010yk} which were shown to essentially allow for any desired background expansion history. Furthermore, they also include the mimetic cubic Galileon which can, for instance, reproduce the expansion history of the LCDM model by a suitable choice of $b(\varphi)$ \cite{Arroja:2015wpa}.
In this case, $f_8=0$ and $f_7=-M_{Pl}^2=-f_9=f_{10}/2$. This implies that there is no effective anisotropic stress, i.e. $\Phi=\Psi$ (because we assume $\Pi_\mathfrak{f}=0$ for all matter species in this paper) and the coupling with gravity is the standard one.
Eq. (\ref{mc}) then becomes
\begin{equation}
\Phi'+\mathcal{H}\Phi+\left(-\frac{a^2\sum_{\mathfrak{f}=r,b}(\bar{\rho}_\mathfrak{f}+\bar{p}_\mathfrak{f})}{2M_{Pl}^2}+\mathcal{H}^2-\mathcal{H}'\right)v_\varphi
+\frac{a^2}{2M_{Pl}^2}\sum_{\mathfrak{f}=r,b}(\bar{\rho}_\mathfrak{f}+\bar{p}_\mathfrak{f})v_\mathfrak{f}=0.
\label{mc2}
\end{equation}
We can see that this mimetic model effectively introduces a fluid which clusters on all scales (i.e. $c_s=0$).
In order to find the solution for the Newtonian (Bardeen) potential $\Phi$ one only needs to specify one function, the background expansion rate $\mathcal{H}$. $\mathcal{H}$ can be found by solving the background equations of motion once the functions $b$, $K$ and $G_3$ are specified. However, here we follow a different approach. We assume that a solution for the inverse problem exists, that is,  we assume that suitable choices for those functions exist such that we can have for example the background expansion history of a CDM plus perfect fluid DE model in GR (hereafter called perfect fluid dark energy or PFDE).
Given the freedom that these models allow for the background history, for future work, it would be interesting to reconstruct the background from observations and then use that background in Eq. (\ref{mc2}) to find the prediction for the potential $\Phi$. At the same time, it would also be interesting to solve the inverse problem to find the constraints that the free functions $b$, $K$ and $G_3$ should obey.

The background expansion of a PFDE model is given by
\begin{equation}
2M_{Pl}^2(\mathcal{H}^2-\mathcal{H}')=a^2\sum_{\mathfrak{f}=r,m,DE}(\bar{\rho}_\mathfrak{f}+\bar{p}_\mathfrak{f}),
\end{equation}
where here $r=\gamma+\nu$ and $m=b+CDM$.
Plugging this equation in Eq. (\ref{mc2}) we get
\begin{equation}
\Phi'+\mathcal{H}\Phi+\frac{a^2}{2M_{Pl}^2}\left[(\bar{\rho}_{CDM}+\bar{p}_{CDM})v_\varphi+(\bar{\rho}_{DE}+\bar{p}_{DE})v_\varphi+\sum_{\mathfrak{f}=r,b}(\bar{\rho}_\mathfrak{f}+\bar{p}_\mathfrak{f})v_\mathfrak{f}\right]=0.\label{mc3}
\end{equation}
Recalling that $v_\varphi$ behaves as a dust velocity following Eq. (\ref{dustv}), one can compare this equation with the second equation in (\ref{EEPF}) to find that they are equal if $v_{DE}$ in (\ref{EEPF}) behaves as a dust velocity. That is guaranteed to be the case if $c_s\equiv c_{(DE)s}=0$ by using Eq. (\ref{vp}) (and if $\Pi_{DE}=0$, i.e. for PFDE).
This shows that for the simpler models defined in Eq. (\ref{models}), if all matter species have $\Pi=0$ and if the background expansion is identical to the background expansion of a perfect fluid DE in Einstein's gravity with any EOS, then the mimetic models predict exactly the same solution for $\Phi$ (which is equal to $\Psi$) as in the perfect fluid DE model with $c_s=0$. This result also implies that the mimetic model with the LCDM background history also gives the same prediction for $\Phi$ as the LCDM model. This is the main result of this section and one of the main results of this paper.

The previous equivalence argument assumes adiabatic initial conditions. One important difference arises in the presence of non-adiabatic initial conditions. The perfect fluid DE model with $c_s=0$ can support non-adiabatic initial conditions between the CDM and DE velocities while in the mimetic model this situation cannot be realized. This could in principle be used to distinguish these otherwise equivalent models at the level of linear perturbations (assuming the mimetic model has the background of a perfect fluid DE model).

If one is interested in the late time universe, well after the time when radiation became negligible, then one can combine Eqs. (\ref{dustv}) and (\ref{mc3}), assuming initial conditions such that $v_\varphi=v_b$, to find
\begin{equation}
\Phi''+\Phi'\left(3\mathcal{H}+\tilde{\Gamma}\right)+\Phi\left(\mathcal{H}^2+2\mathcal{H}'+\tilde{\Gamma}\mathcal{H}\right)=0,
\end{equation}
where $\tilde{\Gamma}$ is defined as
\begin{equation}
\tilde\Gamma\equiv\frac{-\mathcal{H}''+\mathcal{H}\mathcal{H}'+\mathcal{H}^3}{\mathcal{H}'-\mathcal{H}^2}.\label{tildeGamma}
\end{equation}
In the LCDM model and at the late time, $\tilde{\Gamma}=0$, as shown in Appendix \ref{app:evoleq}.
The previous equation can be solved analytically \cite{Lim:2010yk} (see also \cite{Arroja:2015yvd}) and one finds
\begin{equation}
\Phi_2\propto1-\frac{H}{a}\int \frac{da}{H},
\end{equation}
for the growing mode (same form of solution as Eq. (\ref{phicsol})) and $\Phi_1\propto H/a$, for the decaying mode. It is worth mentioning that if one considers observations of $\Phi$ only, in a cosmological background, then the present model, given by Eq. (\ref{models}), is observationally indistinguishable from the unified dark matter and dark energy model of Ref. \cite{Lim:2010yk}. If one considers other observables in other non-cosmological backgrounds it may be possible to distinguish these two models, however that investigation is beyond the scope of the present paper.

\section{Numerical analysis\label{sec:numerical}}

In this paper, we use two different numerical codes to compute the power spectrum for the LCDM model as well as for the PFDE and mimetic models.
One is the fitting formula of the transfer function originally developed by Eisenstein and Hu (EH) \cite{Eisenstein:1997ik} for the LCDM model and further extended by \cite{Hu:2001fb,Takada:2006xs}, denoted here by EHT, for the PFDE model. In Appendix \ref{app:PSPFDE} we briefly review these fitting-function results. Another one is the code, developed by us, which directly integrates the evolution equations of mimetic models and perfect fluid models for DE, as briefly described in the next subsection.

In the second subsection, we describe the N-body simulations from which the power spectra of matter and galaxies were extracted and compared to our numerical results.  In Section \ref{sec:distinguish}, we used these measurements as a proxy for future observations. In the third subsection, we compare the power spectra obtained by the two codes with the exact solution based on the Boltzmann code for the LCDM model \cite{Lewis:1999bs}. We also discuss the accuracy of the two codes by comparing to each other for the PFDE model.

\subsection{Description of our numerical hydrodynamical code \label{subsec:code}}

We solve numerically Eqs. (\ref{dustv}), (\ref{mc2}), (\ref{deltap}) and (\ref{vp}) for the mimetic model, where the perfect fluids considered include only baryons and radiation. For the PFDE models we solve numerically Eqs. (\ref{deltap}), (\ref{vp}) and (\ref{EEPF}), where in this case the perfect fluids include radiation, baryons, cold dark matter and dark energy with a CPL equation of state (gravity is assumed to be described by GR). The LCDM model is a particular case of the PFDE models where $w_{DE}=-1$ and there are no perturbations in the dark energy fluid. For the initial conditions, set during the radiation era, we choose adiabatic initial conditions ($\delta\rho/\bar{\rho}'=\delta\rho_\mathfrak{f}/\bar{\rho}_\mathfrak{f}'$ at the initial time and where $\mathfrak{f}$ denotes the different species) as
\begin{equation}
\delta_{ri}=-2\Phi_i,\quad
\delta_{mi}=-\frac{3}{2}\Phi_i,\quad
\delta_{DEi}=-\frac{3}{2}(1+w_{DEi})\Phi_i,\quad
\tilde{v}_i=-\frac{1}{2}\Phi_i=\tilde{v}_{ri}=\tilde{v}_{mi}=\tilde{v}_{DEi},
\end{equation}
where the subscript $i$ denotes at the initial time $N_i$. We use the number of efolds, $N=\ln a$, as the time variable. $\Phi_i$ is the initial value of the potential. We use that $|\mathcal{R}_i|\approx3/2|\Phi_i|$ on large scales and Eq. (\ref{primordialspectrum}) to correctly normalize $\Phi_i$.
For the initial condition for $\tilde{v}_\varphi$ in the mimetic models, we choose $\tilde{v}_\varphi=-\frac{1}{2}\Phi_i=\tilde{v}_{ri}=\tilde{v}_{bi}$.

We start the numerical code during the radiation era, at $N_i=-15$ (we choose that the scale factor today is $a_0=1$) i.e. initial redshift $z_i$ of about $z_i\sim3\times10^6$. The code has two versions, one where we neglect the fact that until recombination baryons and photons are strongly coupled, and another version where we use a simple toy model for this coupling. We assume that recombination happens instantaneously at redshift $z_d$, so that after this redshift, photons and baryons are totally decoupled. Before this redshift we take the coupling into account by setting the baryon velocity equal to the photon velocity, i.e. $v_b=v_\gamma$. This toy model for recombination follows a similar logic to the one in Chapter 6 of \cite{Weinberg:2008zzc}. We will characterize its accuracy in the next subsection.
In this case, we assume that the massless neutrinos and antineutrinos contribute an energy density corresponding to three species at a temperature of $(4/11)^{1/3}$ of the temperature of the photons, and one has
\begin{eqnarray}
\Omega_{\gamma 0}=\frac{1}{1+3\times\frac{7}{8}\left(\frac{4}{11}\right)^{4/3}}\Omega_{r0},\qquad
\Omega_{\nu 0}=\frac{3\times\frac{7}{8}\left(\frac{4}{11}\right)^{4/3}}{1+3\times\frac{7}{8}\left(\frac{4}{11}\right)^{4/3}}\Omega_{r0}.
\end{eqnarray}

For the value of $z_d$ we use Eq. (4) of \cite{Eisenstein:1997ik}, which is a fit to numerical recombination results. It reads
\begin{eqnarray}
z_d&=&1291\frac{\left(\Omega_{m0}h^2\right)^{0.251}}{1+0.659\left(\Omega_{m0}h^2\right)^{0.828}}\left[1+b_1\left(\Omega_{b0}h^2\right)^{b_2}\right],
\\
b_1&=&0.313\left(\Omega_{m0}h^2\right)^{-0.419}\left[1+0.607\left(\Omega_{m0}h^2\right)^{0.674}\right],
\quad
b_2=0.238\left(\Omega_{m0}h^2\right)^{0.223}.
\end{eqnarray}

\subsection{N-body simulations and power spectra\label{subsec:Nsims}}

In order to discuss if one can distinguish the mimetic models from the LCDM model and the PFDE models, we use measurements of the power spectra from LCDM N-body simulations as a proxy for future observations. For this purpose we use the power spectra of matter, $P_m(k)$, and galaxies, $P_g(k)$, measured in \cite{Okumura:2011pb} and \cite{Okumura:2012xh}, respectively. The simulations were run by \cite{Desjacques:2008vf} assuming the following cosmological parameters in a spatially flat universe: $\Delta_\mathcal{R}^2(k_0=0.02\mathrm{Mpc}^{-1})=2.21\times10^{-9}$, $n_s=0.96$, $h=0.7$, where $h\equiv H_0/(100\mathrm{Km}\, \mathrm{s}^{-1} \mathrm{Mpc}^{-1}$), $\Omega_{m0}=0.279$, $\Omega_{b0}=0.0462$ and $\Omega_{r0}=8.48\times10^{-5}$, which corresponds to a CMB temperature of $T_0=2.725K$ for our value of $h$.
$1024^3$ particles of mass $3.0\times 10^{11}h^{-1}M_\odot$ are employed in a cubic box of side $1600h^{-1}$Mpc at the redshift of a typical galaxy redshift survey, $z=0.509$.
12 independent realizations are used to reduce the statistical scatters, which corresponds to the effective volume of $\sim 50~(h^{-1}{\rm Gpc})^3$, relevant to the EUCLID survey. The error bars are estimated by the dispersion among realizations, and in the following analysis we show the error of the mean, which is the dispersion divided by $\sqrt{12}.$ The error bars do not include contributions from observational systematics, and we thus regard these realizations as an ideal observation.

We can simply use the measurement of the power spectrum and convert it to $P_\Phi$ by using the Poisson equation in LCDM. However, the fractional error of the matter power, $(\Delta P_m/P_m)$, is known to be much smaller than that of the galaxies due to the large number of dark matter particles. In order to make a conservative comparison, we also use the power spectrum and its error taken from the galaxy power spectrum assuming the linear bias, $P_m = P_g/b^2$ and $\Delta P_m = \Delta P_g/b^2$, respectively. The power spectrum was computed from the mock galaxy catalogs relevant to the SDSS BOSS, where $b\simeq 2.16$ (see \cite{Okumura:2012xh} for details).

\subsection{Comparison between the results of the hydrodynamical code and fitting functions\label{sec:codeacc}}

In order to compute model predictions for the power spectra, for consistency we assume the same cosmological parameter set as that used to run the simulations.
They are close to the WMAP cosmological parameters.
We believe that our conclusions should not depend strongly on this choice.

\begin{figure}[bt]
\adjustbox{trim={.00\width} {.05\height} {0.00\width} {.15\height},clip}
  {\includegraphics[scale=0.8]{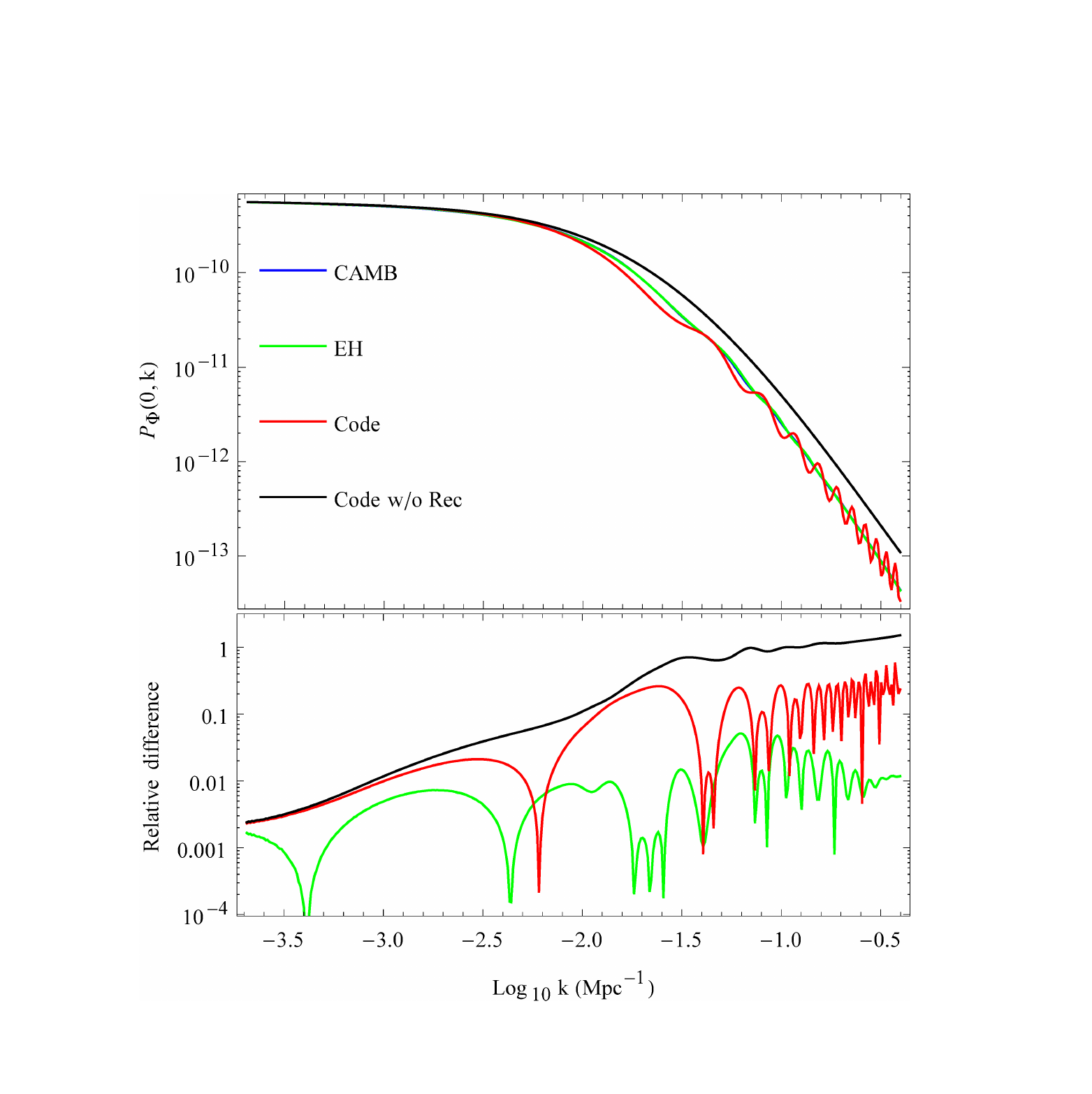}}
\caption{Top: Plot of $P_\Phi$ today versus wavenumber k for the LCDM model.
The blue line is the result obtained with CAMB (practically indistinguishable from the green line). The green line is the fitting function of Eisenstein and Hu. The result of our numerical code neglecting the coupling between photons and baryons before recombination is the black line. In the red line the coupling is taken in account using a simple model as described in the previous subsection. Bottom: Plot of the absolute value of the relative differences with respect to the CAMB result. It uses the same color code. E.g. the red line is the relative difference between CAMB and the result of our code with recombination taken into account.}
\label{Fig:testcode}
\end{figure}

In Fig. \ref{Fig:testcode}, we plot the power spectrum of $\Phi$, $P_\Phi$ today, versus wavenumber $k$ for the LCDM model. We also plot the absolute value of the relative difference with respect to the result of the CAMB Boltzmann code \cite{Lewis:1999bs}.
One can see that the fitting function result of Eisenstein and Hu, Eq. (\ref{EH}), agrees with the exact numerical result obtained from CAMB to better than about $5\%$. When the coupling between baryons and photons before recombination is ignored, we see that the result deviates from the true one by about $100\%$ on small scales, $k\lesssim10^{-1.5}\mathrm{Mpc}^{-1}$. This assumption has been often made in the literature without such an accuracy test.
Our code takes into account the coupling with a very simple model and it improves the accuracy to about $30\%$ for $10^{-2}\mathrm{Mpc}^{-1} \lesssim k\lesssim 10^{-1}\mathrm{Mpc}^{-1}$. For $k\lesssim 10^{-2}\mathrm{Mpc}^{-1}$ our results agree to about $5\%$ with CAMB. Taking the coupling into account reduces the power for scales smaller than about $10^{-2}\mathrm{Mpc}^{-1}$. This is a well-known effect for two decades \cite{Eisenstein:1997ik}. It also introduces baryon acoustic oscillations on these scales (compare the black and red curves). However the simple coupling prescription that we use over amplifies these oscillations with respect to the correct coupling treatment in CAMB. From now on, when we present numerical results obtained with our code, the coupling is always taken into account using the simple coupling model.

\begin{figure}[bt]
\includegraphics[scale=0.8]{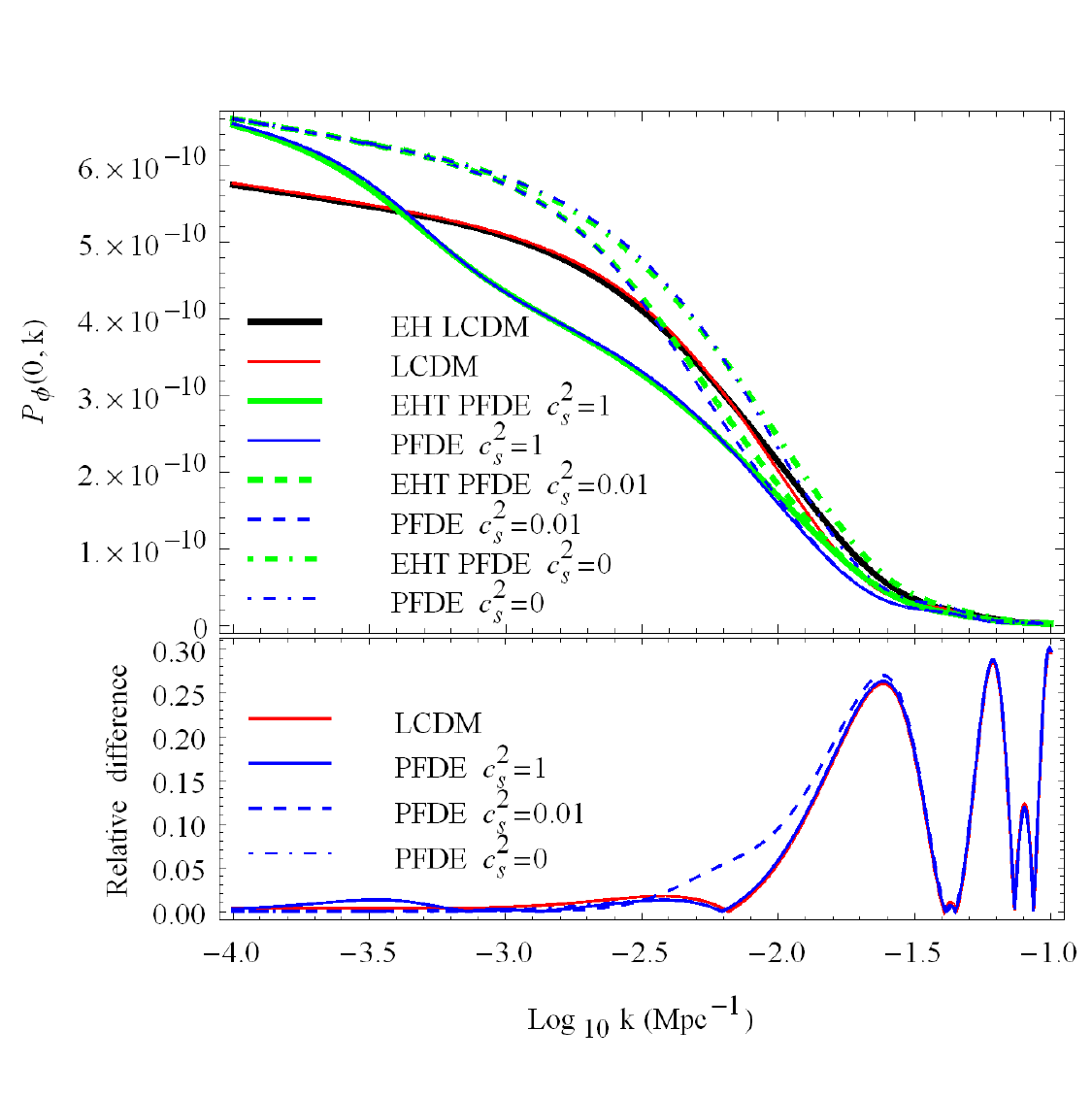}
\caption{Top: Plot of $P_\Phi$ today versus wavenumber k for the LCDM model and three PFDE models with different sound speeds. The EOS of the PFDE used to produce this plot was $w_0=-0.7,\,w_a=0$.
Both black and red lines are for the LCDM model. The black line is the result using the fitting functions of EH, while the red line is the result of our code.
The green lines are the results for the PFDE model obtained with the fitting functions and some numerical integration following EHT for three values of the sound speed. The blue lines where obtained with our code. Bottom: Plots of the absolute value of the relative differences between the four pairs of lines, LCDM, PFDE $c_s^2=1$,  PFDE $c_s^2=0.01$ and  PFDE $c_s^2=0$.
}
\label{Fig:testcodePFDE}
\end{figure}

In Fig. \ref{Fig:testcodePFDE}, we plot the power spectrum of $\Phi$ today versus wavenumber k for the PFDE model with three different values of the sound speed and EOS $w_{DE}=w_0=-0.7$ and for the LCDM model. For each model and parameter combination we plot the results of our code and the results obtained with the fitting functions of EHT (Eq. (\ref{EHT})). We also plot the absolute value of the relative difference between our code results and the fitting functions. On the largest scales, the difference between the amplitudes of the power spectrum between the LCDM and PFDE models is due to the presence of DE perturbations which can be this large due to the fact the EOS is significantly different from $-1$. For smaller sound speed, the sound horizon of the DE perturbations becomes smaller, which implies that these perturbations can cluster on smaller scales, and this explains why the model with $c_s^2=0$ has the highest power, followed by the line with $c_s^2=0.01$ and finally the least power for the line $c_s^2=1$.
Ref. \cite{Hu:2001fb} showed that the fitting function agrees well (about $10\%$ accuracy) with respect to the numerical result of the full integration of the Einstein plus Boltzmann system of equations.
In the bottom plot, we show the absolute value of the relative differences between the results obtained with our code and the fitting functions. One can see that our code agrees well with the fitting functions results (to about $10\%$) for $k\lesssim 10^{-2}\mathrm{Mpc}^{-1}$. For $k\gtrsim 10^{-2}\mathrm{Mpc}^{-1}$, our results lose accuracy (relative difference of about $30\%$) and this is due to the very simple model of recombination that we use and possibly also due to the hydrodynamical approximation for all the fluids involved. Essentially this is an error of the transfer functions that one would get with our code. We see that this loss of accuracy is similar for all 4 models.

As one can see, our numerical code has advantages and disadvantages. The advantages are that it is transparent, involves simple equations (hydrodynamical equations instead of full Boltzmann equations and a toy model for recombination) and roughly captures the relevant physics. Furthermore, we can use any EOS for the DE and any background expansion history. The disadvantages are that it is not accurate for scales smaller than $k\sim 10^{-2}\mathrm{Mpc}^{-1}$. In fact, in the current era of precision cosmology even the fitting functions that we use here and describe briefly in Appendix \ref{app:PSPFDE} are not sufficiently accurate to do parameter estimation. Nowadays, the use of an accurate Boltzmann-Einstein system integrator, like CAMB, is imperative. However, the goal of this paper is not to do parameter estimation or constrain the parameters of the models. Our goals are more modest and start by showing that these mimetic models, under some parameter choices give reasonable predictions for the linear power spectrum, then we want to show in which circumstances we can distinguish these models from other popular models for the dark universe. For these purposes, our code is enough and it also motivates future studies, like for instance, the implementation of our equations in CAMB and the task of cosmological parameter estimation in mimetic models.

\section{Distinguishability between mimetic and perfect fluid dark energy models\label{sec:distinguish}}

The goal of this section is to discuss the distinguishability between mimetic models (that describe the entire dark universe with a single entity, the mimetic scalar field) and popular models that explain the accelerated expansion, as for instance the LCDM model and perfect fluid models for CDM and DE (assuming gravity is described by GR). With this goal in mind, we compute the power spectrum of $\Phi$ from the power spectra measured from the N-body simulations and we take the error bars from these measurements as an estimate of the statistical error bars that future LSS surveys will obtain. If the differences between the predictions of the different model/parameter combinations are larger than the error bars of the measurements we argue that those models/parameter combinations are distinguishable in the future and a more accurate analysis of parameter estimation would be desirable.
As we showed previously in Section \ref{sec:equiv}, the distinguishability between models is strongly depend on the assumed background expansion history. For instance, we showed that if one fixes the background history to the background history of a PFDE model then the mimetic model predicts that the linear power spectrum of $\Phi$ will be exactly the same as the PFDE model with $c_s=0$ and therefore these two models cannot be distinguished using this observable. In particular, for a LCDM background we also showed that the mimetic model and the LCDM model cannot be distinguished.

\begin{figure}[bt]
\includegraphics[scale=0.8]{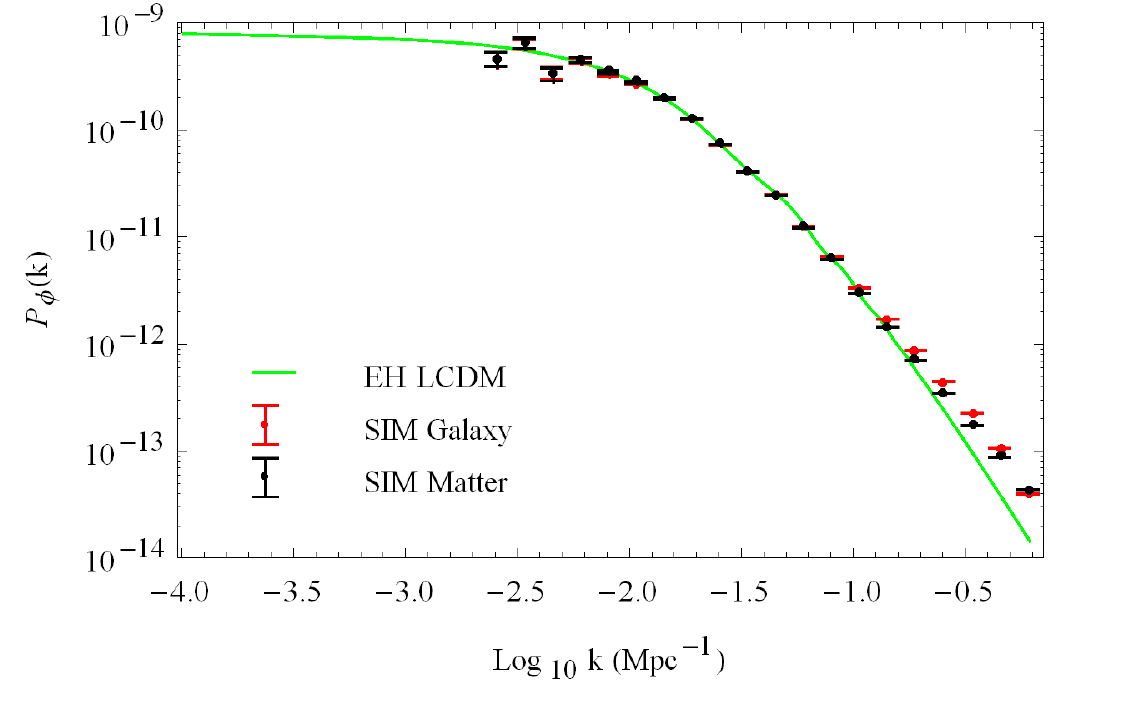}
\caption{Plot of $P_\Phi$ versus wavenumber k for the LCDM model (green line), using measurements of the power spectrum of galaxies (red points) and matter (black points) from N-body simulations. The redshift of the simulations is $z_{sim}=0.509$ and the power spectrum is evaluated at the same redshift.
}
\label{Fig:plotsimLCDM}
\end{figure}

In Fig. \ref{Fig:plotsimLCDM}, one can see that the simulation measurements start to deviate from the EH fitting function for scales smaller than $k\sim10^{-1}\mathrm{Mpc}^{-1}$. We take this as the scale at which non-linear effects, not taken into account in the fitting function, start to become important. Therefore, to draw our following conclusions we ignore scales smaller than this non-linearity scale. On very large scales, $k\lesssim 10^{-3}$, one expects GR effects to come into play (e.g. see \cite{Challinor:2011bk,Jeong:2011as}). These may introduce corrections not taken into account here. However, the scales are so large that they are not relevant for the scales measured in the N-body simulations.

\begin{figure}
\centering
\subfloat{\label{Fig:ComparisonWithLCDMm07}\includegraphics[width=0.5\textwidth]{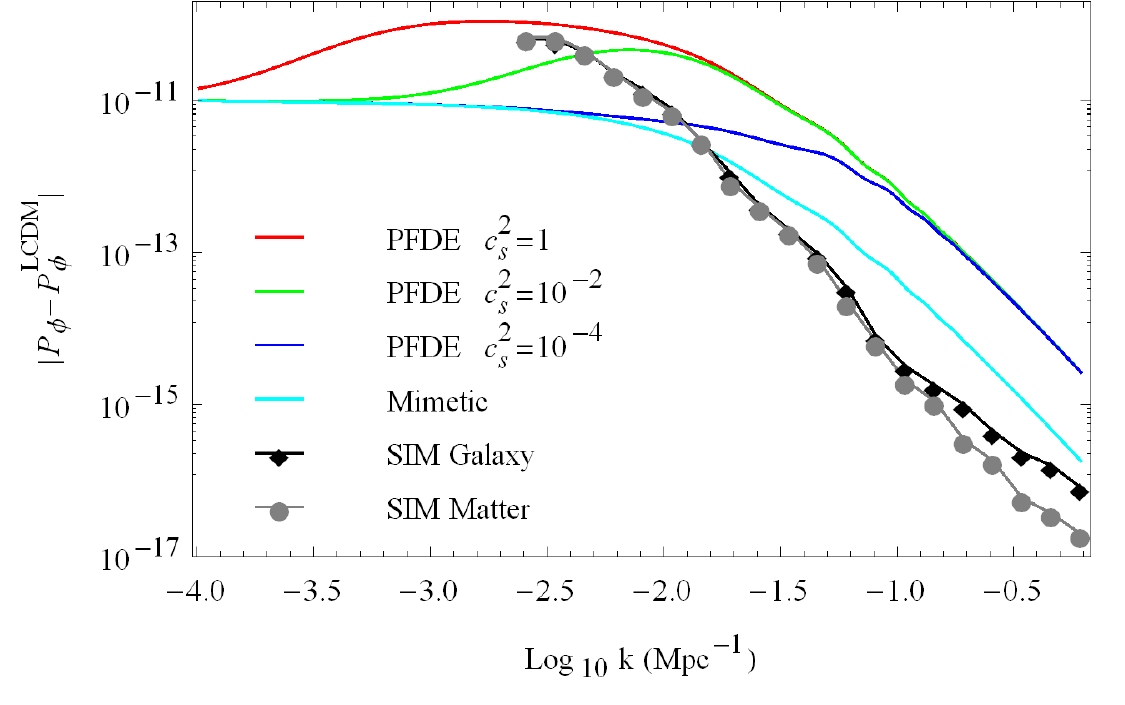}}
\subfloat{\label{Fig:ComparisonWithLCDMm095}\includegraphics[width=0.5\textwidth]{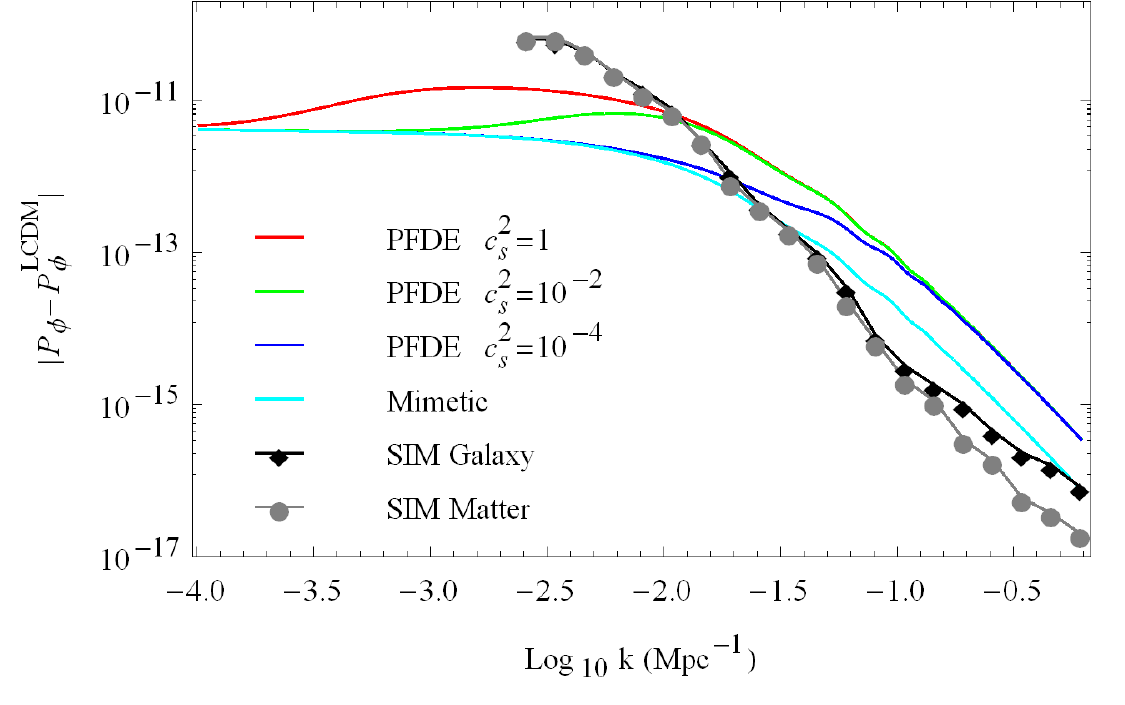}}
\caption{Plot of the absolute value of the difference between the power spectra for the different sound speeds, the mimetic model and the power spectrum of LCDM model (EH fitting function). All spectra are evaluated at $z_{sim}=0.509$ and plotted versus wavenumber k. The joined black diamonds are the size of the error bars from measurements of the power spectrum of galaxies from the LCDM N-body simulations. The joined gray circles are the size of the error bars from measurements of the power spectrum of matter in the same simulations. The EOS of the DE is $w_{DE}=-0.7$ (left) and $w_{DE}=-0.95$ (right).
}
\label{Fig:ComparisonWithLCDM}
\end{figure}

In Fig. \ref{Fig:ComparisonWithLCDM}, we plot the absolute value of the difference of the power spectra for the different sound speeds and the mimetic model (using the EHT fitting function) relative to that of the LCDM model (using the EH fitting function). We choose $w_{DE}=-0.7$ and $w_{DE}=-0.95$ for the plots on the left and right panels, respectively. If $w_{DE}$ is constant in time then from the Planck observations of the CMB \cite{Ade:2015xua} we know that $w_{DE}$ should be close to $-1$. If fact, the value $-0.7$ is already rule-out by these observations (nevertheless it is shown here for illustrative purpose only), while $w_{DE}=-0.95$ is still allowed. One can see that for the mimetic model with the background $w_{DE}=-0.7$ (which gives the same prediction as the PFDE model with zero sound speed), one can distinguish it from LCDM. For the other values of the sound speed the models will be more easily distinguished in the future. However, when the EOS approaches $-1$, as shown in the right panel, we see that the mimetic model will hardly be distinguishable from LCDM. However for the other values of the sound speed there is still a possibility to distinguish those models.
From the figure one can also see that the size of the error bars of the power spectrum $P_\Phi^{LCDM}$ obtained from measurements of the galaxy (black diamonds) and matter (gray circles) power spectra only start to differ appreciably for scales smaller than $k\sim10^{-1}\mathrm{Mpc}^{-1}$. Because we exclude these non-linear scales from our discussion, our conclusions are independent of the choice of galaxy or matter errors bars.

\begin{figure}
\centering
\subfloat{\label{Fig:ComparisonWithPFDE1m07}\includegraphics[width=0.5\textwidth]{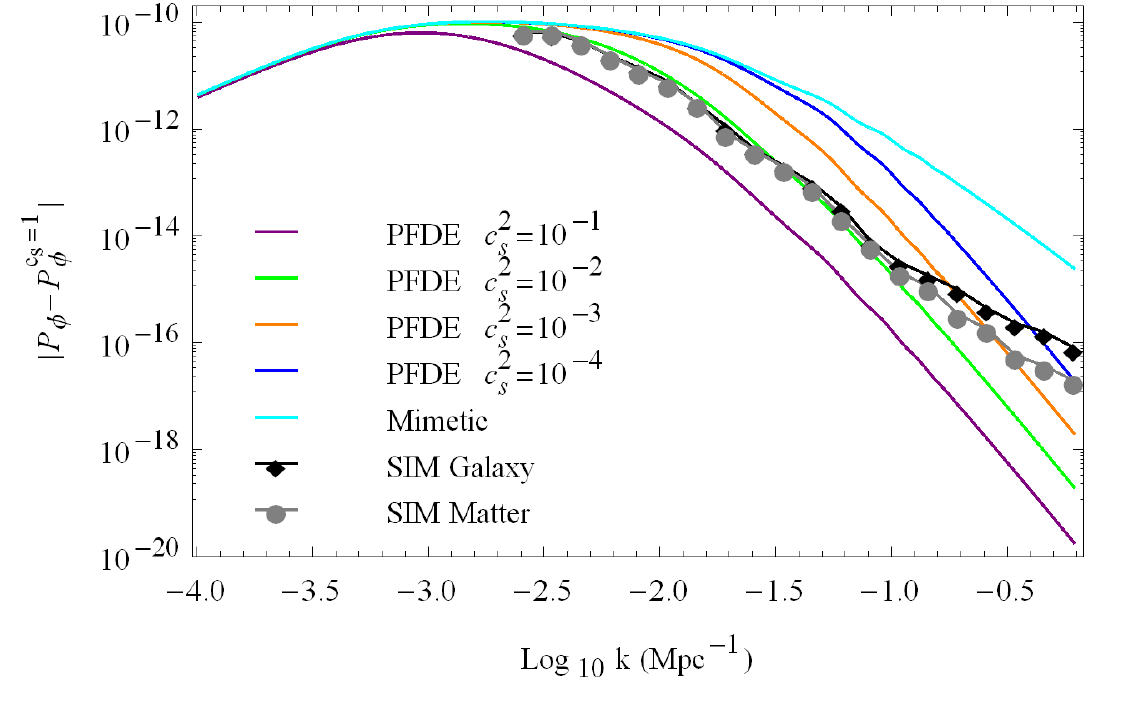}}
\subfloat{\label{Fig:ComparisonWithPFDE1m095}\includegraphics[width=0.5\textwidth]{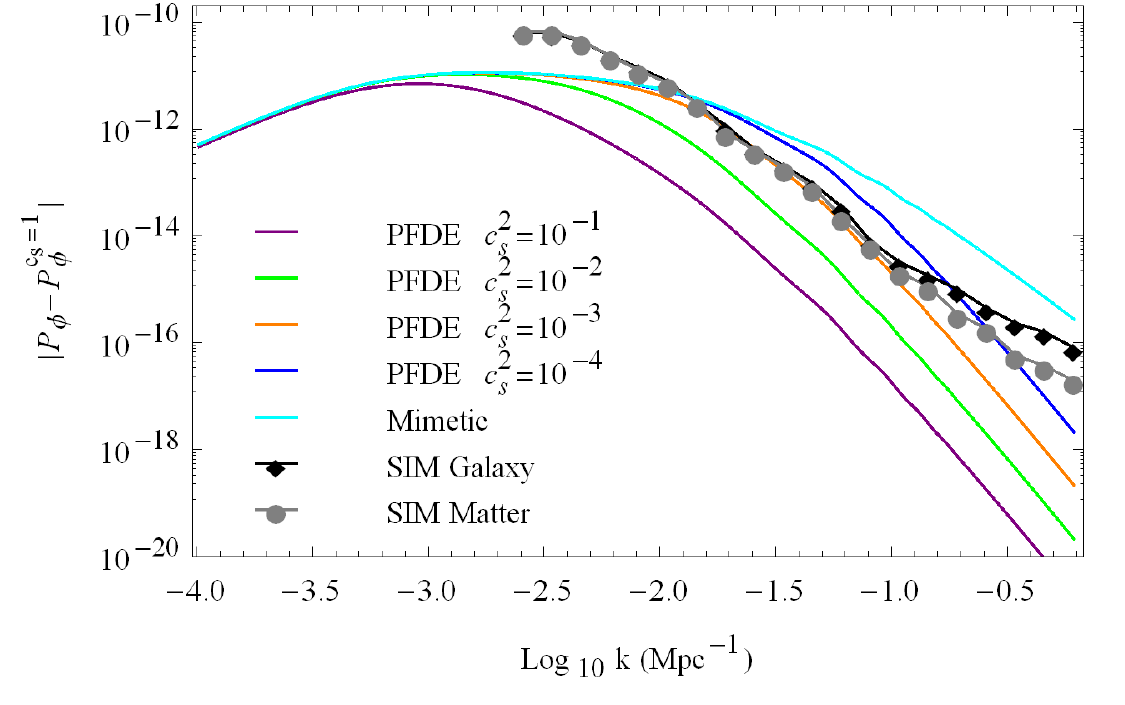}}
\caption{Same as Fig. \ref{Fig:ComparisonWithLCDM} but here the difference is taken with respect to the power spectrum of the PFDE model with $c_s=1$. Here we use the EHT fitting function.
}
\label{Fig:ComparisonWithPFDE1}
\end{figure}

Fig. \ref{Fig:ComparisonWithPFDE1} is a similar plot to Fig. \ref{Fig:ComparisonWithLCDM}, but here show the differences relative to the power spectrum for $c_s=1$ of the PFDE model (all using the EHT fitting function). Again, for the plots on the left and right panels, we adopt $w_{DE}=-0.7$ and $w_{DE}=-0.95$, respectively. One can see that with future data one might be able to marginally distinguish models with $c_s^2\lesssim10^{-2}$ from the model with $c_s=1$ for $w_{DE}=-0.7$. This result qualitatively agrees with the findings of \cite{Takada:2006xs}. If $w_{DE}=-0.95$, we will only be able to distinguish models with the sound speed smaller than about $10^{-2}$ from the model with unit sound speed. In particular, we will be able to distinguish the mimetic model from the PFDE model with $c_s=1$, which is one of the popular models of dark energy.

\begin{figure}
\centering
\subfloat{\label{Fig:ComparisonWithLCDMm07m03}\includegraphics[width=0.5\textwidth]{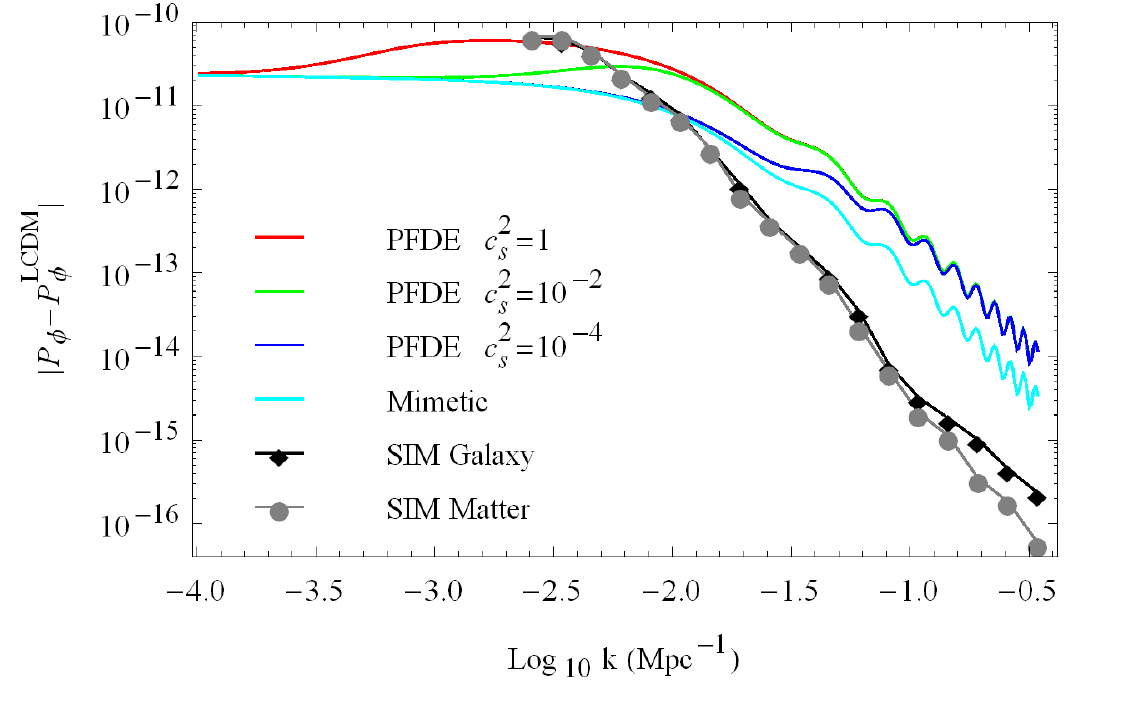}}
\subfloat{\label{Fig:ComparisonWithPFDE1m07m03}\includegraphics[width=0.5\textwidth]{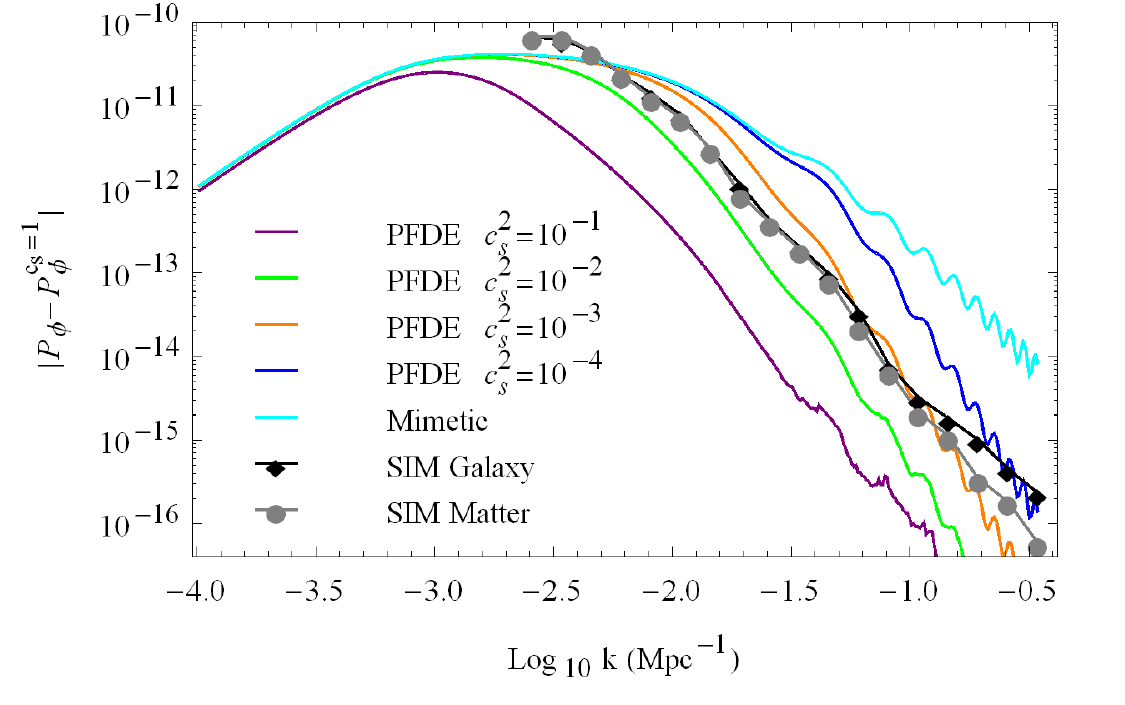}}
\caption{
The left and right panels are the same as in Figs. \ref{Fig:ComparisonWithLCDM} and \ref{Fig:ComparisonWithPFDE1}, respectively, but here we assume a time-dependent EOS as $w_{DE}=-0.7-0.3(1-a)$. These plots use results obtained with our numerical hydrodynamical code.
}
\label{Fig:ComparisonWithLCDMandPFDE1m07m03}
\end{figure}

In Fig. \ref{Fig:ComparisonWithLCDMandPFDE1m07m03}, we do similar comparisons as in the previous figures but this time we adopt a time dependent EOS as $w_{DE}=-0.7-0.3(1-a)$. These values for $w_0$ and $w_a$ are still in the allowed range by Planck observations \cite{Ade:2015rim}. In this figure we use the numerical results from our hydrodynamical code, which has about $30\%$ uncertainty for scales smaller than $k\sim10^{-2}\mathrm{Mpc}^{-1}$. This means that in drawing conclusions from those scales one should allow the models to be about half an order of magnitude above the simulation error bars in order to claim that the models can be distinguished observationally. One finds, that the mimetic models can be distinguished from both LCDM and from  the PFDE model with $c_s=1$. For this background choice, one may be able to discriminate between $c_s=10^{-2}$ (or smaller) from $c_s=1$ for PFDE models.

\section{Conclusion\label{sec:con}}

In this paper, we have proposed to use the mimetic Horndeski gravity as a model for the dark universe, i.e. we use the model as a unified DM and DE model. In this scenario, gravity is modified and a single entity, the mimetic scalar field, describes the phenomena usually associated with DM and DE. We showed that the mimetic component behaves effectively like a perfect fluid with $c_s=0$, and therefore clusters on all scales. We then, for simplicity, considered the mimetic cubic Horndeski model. This model is still quite general and includes various simpler models in the literature, in particular the original mimetic DM model \cite{Chamseddine:2013kea}. It is an interesting example also because it is known to allow almost any desired background expansion history. If one assumes that the background is given by CDM plus perfect fluid DE described by GR with any equation of state, then we showed that the mimetic model predicts exactly the same solution for the Newtonian (Bardeen) potential $\Phi$ at linear order as the PFDE model with the sound speed equal to zero. Therefore the two models are indistinguishable using that observable. Because we assume that the mimetic field and the DM and DE fluids interact with the particles of the Standard model only gravitationally, observables related to gravitational effects are the only relevant ones. A corollary of the previous result is that the mimetic model with the LCDM background history will also give the same prediction for $\Phi$ as the LCDM model. Considering observations of $\Phi$, the cubic mimetic Horndeski model is also indistinguishable, in a cosmological background, from the unified dark matter and dark energy model of Ref. \cite{Lim:2010yk}. The previous results assume adiabatic initial conditions. If the perturbations are not adiabatic one may be able to distinguish between the mimetic and the PFDE models.

We then develop a simple code where we implement the perturbed hydrodynamical equations and the relevant metric equations of motion in the mimetic model that we solve numerically. We use a simple toy model for recombination and thus for small scales our code is not very accurate. Despite of this poor accuracy it is enough to achieve the goals of this paper, which are to show that for some parameter choices one can have reasonable predictions for the linear power spectrum and roughly under which circumstances we can distinguish these mimetic models from other popular DE models. We leave for future work the implementation of the equations that we obtained in well-established Boltzmann integrator codes and after that the task of parameter estimation in mimetic models. Given the present work, this is a well motivated project now.
When the EOS does not depend on time, we made extensive use of the well-known fitting functions for the power spectrum, which are more accurate than our hydrodynamical code.

Using power spectrum measurements from LCDM N-body numerical simulations as a proxy for constraints that future LSS surveys may obtain, we discussed the distinguishability of the mimetic models from LCDM and PFDE models. This distinguishability is strongly dependent on the background to be considered as we showed.
We found that if the mimetic model has the background given by $w_{DE}=-0.7$ then we can distinguish it from the LCDM model. However, it is known that this value of EOS is already rule-out observationally. If the EOS approaches, -1, e.g. it is $w_{DE}=-0.95$, then the two models will hardly be distinguished as expected in light of our new results. For PFDE models with $c_s\neq0$ there is hope to be able to observationally distinguish them from LCDM. Even if the mimetic background is $w_{DE}=-0.95$ we found that we will be able to distinguish the mimetic model from the PFDE model with unity sound speed, which is one of the popular models of DE. If we allow, a time dependent EOS, following the CPL parametrization as $w_{DE}=-0.7-0.3(1-a)$ for example, then we found that the mimetic model can be distinguished from both LCDM and PFDE models with $c_s=1$.

In the future, given that these mimetic models can accommodate almost any background expansion history, instead of fixing the background to be the same as some popular models of DE, it would be interesting to reconstruct   the background from observations and then use it to compute the linear predictions. This is beyond the scope of this paper.
The investigation into the non-linear regime and the potential issue of caustic appearance is also left for future work.
Finally, as far as we know, the issue of a consistent linear galaxy bias treatment in clustering DE models, including mimetic models, is an open problem which we are planning to address next. This is a necessary step before one can make use of LSS constraints for the growth factor for example.

\emph{Note added:} After this paper was published in the arXiv, Refs. \cite{Takahashi:2017pje,Langlois:2018jdg} appeared and they argued that in the presence of external matter, modelled by a k-essence scalar field, the models discussed in this work (and many other mimetic models present in the literature) contain either a ghost or a gradient instability. Currently we are investigating whether or not their findings can be generalized to the models discussed here, where the external matter is assumed to be described by fluids. It is worth mentioning that there are mimetic models (see the model discussed in Ref. \cite{Ramazanov:2016xhp} for instance) where the presence of a ghost instability is no cause of concern even quantum mechanically because the timescale of the vacuum instability can be made sufficiently long to render the model phenomenologically viable \cite{Ramazanov:2016xhp}.
\section{Acknowledgments}

FA would like to thank Fabien Nugier for useful discussions. TO is grateful to Masahiro Takada for useful conversations. Some computations in this paper were performed using Mathematica\footnote{https://www.wolfram.com/mathematica/}. FA is supported by the National Taiwan University (NTU) under Project No. 103R4000 and by the NTU Leung Center for Cosmology and Particle Astrophysics (LeCosPA) under Project No. FI121. TO acknowledges support from the Ministry of Science and Technology of Taiwan under the grant MOST 106-2119-M-001-031-MY3. NB and SM acknowledge partial financial support by
the ASI/INAF Agreement I/072/09/0 for the Planck LFI Activity of Phase E2.

\appendix

\section{A second order evolution equation for $\Phi$ in mimetic Horndeski gravity coupled to a fluid\label{app:evoleq}}

In this appendix we obtain a second order differential equation for the Newtonian (Bardeen) potential in general mimetic Horndeski models plus a fluid generalizing Eq. (47) of \cite{Arroja:2015yvd}. We then particularize the result for the simpler mimetic models defined in Eq. (\ref{models}).

Using the background equations (\ref{eq0}) and the first order equations (\ref{eq1})-(\ref{eq6}), for a general function $b(\varphi)$, one can manipulate them to obtain an evolution equation for $\Phi$ as
\begin{eqnarray}
&&B_3\Phi''+\Phi'\left[B_2+B_3'+B_3\left(\frac{\bar{\varphi}'\bar{b}_{,\varphi}}{2\bar{b}}-\frac{B_1'}{B_1}\right)\right]
+\Phi\left[B_1\bar{\varphi}'+B_2'+B_2\left(\frac{\bar{\varphi}'\bar{b}_{,\varphi}}{2\bar{b}}-\frac{B_1'}{B_1}\right)\right]
\nonumber\\
&&
+B_5\Pi''
+\Pi'\left[B_4+B_5'+B_5\left(\frac{\bar{\varphi}'\bar{b}_{,\varphi}}{2\bar{b}}-\frac{B_1'}{B_1}\right)\right]
+\Pi\left[B_4'+B_4\left(\frac{\bar{\varphi}'\bar{b}_{,\varphi}}{2\bar{b}}-\frac{B_1'}{B_1}\right)\right]
\nonumber\\
&&
-\left[a^2\left(\bar{\rho}+\bar{p}\right)v\right]'
-a^2\left(\bar{\rho}+\bar{p}\right)v\left(\frac{\bar{\varphi}'\bar{b}_{,\varphi}}{2\bar{b}}-\frac{B_1'}{B_1}\right)=0,
\end{eqnarray}
where $B_1$, $B_2$ and $B_3$ are defined as
\begin{eqnarray}
B_1&=&\frac{f_{10}f_8}{f_7}\left(\frac{f_7'}{f_7}-\frac{f_8'}{f_8}+\frac{\bar{\varphi}'\bar{b}_{,\varphi}}{2\bar{b}}\right)+f_{20}-\frac{\bar{\varphi}'\bar{b}_{,\varphi}}{2\bar{b}}f_{11}+a^2\frac{\bar{E}+\bar{T}}{\bar{\varphi}'},
\nonumber\\
B_2&=&\frac{f_{10}f_9}{f_7}\left(\frac{f_7'}{f_7}-\frac{f_9'}{f_9}\right)+f_{14}+\bar{\varphi}'\left(f_{11}-\frac{f_{10}f_8}{f_7}\right),
\quad
B_3=\frac{2f_9^2}{f_7},
\quad
B_4=a^2\frac{f_{10}}{f_7}\left(\frac{f_7'}{f_7}-2\mathcal{H}\right),
\quad
B_5=-a^2\frac{f_{10}}{f_7}.\nonumber\\
\end{eqnarray}

For the models (\ref{models}), the previous evolution equation can be written as
\begin{equation}
C_1\Phi+C_2\Phi'+\Phi''+C_3\Pi+C_4\Pi'+C_5\Pi''+C_6a^2\left(\bar{\rho}+\bar{p}\right)v-\frac{a^2}{2M_{Pl}^2}\left(\delta p+\frac{2}{3}\partial^2\Pi\right)=0,\label{eomSMM}
\end{equation}
where
\begin{eqnarray}
C_1&=&2\mathcal{H}'-\mathcal{H}\frac{\left[a^2\left(\bar{\rho}+\bar{p}\right)\right]'-2M_{Pl}^2\left(\mathcal{H}'-\mathcal{H}^2\right)\left(\tilde\Gamma+\mathcal{H}\right)}{a^2\left(\bar{\rho}+\bar{p}\right)+2M_{Pl}^2\left(\mathcal{H}'-\mathcal{H}^2\right)},
\quad
C_2=\frac{-a^2\left(\bar{\rho}+\bar{p}\right)'+2M_{Pl}^2\left(\mathcal{H}'-\mathcal{H}^2\right)\left(\tilde\Gamma+3\mathcal{H}\right)}{a^2\left(\bar{\rho}+\bar{p}\right)+2M_{Pl}^2\left(\mathcal{H}'-\mathcal{H}^2\right)},
\nonumber\\
C_3&=&\frac{2a^2}{M_{Pl}^2}\left[\mathcal{H}'-\mathcal{H}\frac{a^2\left(\bar{\rho}+\bar{p}\right)'-a^2\mathcal{H}\left(\bar{\rho}+\bar{p}\right)-2M_{Pl}^2\left(\mathcal{H}'-\mathcal{H}^2\right)\left(\tilde\Gamma+4\mathcal{H}\right)}{a^2\left(\bar{\rho}+\bar{p}\right)+2M_{Pl}^2\left(\mathcal{H}'-\mathcal{H}^2\right)}\right],
\nonumber\\
C_4&=&-\frac{a^2}{2M_{Pl}^2}\frac{2a^2\left(\bar{\rho}+\bar{p}\right)'-6a^2\mathcal{H}\left(\bar{\rho}+\bar{p}\right)-4M_{Pl}^2\left(\mathcal{H}'-\mathcal{H}^2\right)\left(\tilde\Gamma+6\mathcal{H}\right)}{a^2\left(\bar{\rho}+\bar{p}\right)+2M_{Pl}^2\left(\mathcal{H}'-\mathcal{H}^2\right)},
\quad
C_5=\frac{a^2}{M_{Pl}^2},
\nonumber\\
C_6&=&\frac{1}{2M_{Pl}^2}\frac{-a^2\bar{p}'+2M_{Pl}^2\left(\mathcal{H}'-\mathcal{H}^2\right)\tilde\Gamma}{a^2\left(\bar{\rho}+\bar{p}\right)+2M_{Pl}^2\left(\mathcal{H}'-\mathcal{H}^2\right)},
\end{eqnarray}
and $\tilde\Gamma$ is defined as in Eq. (\ref{tildeGamma}).

The previous equation is the generalization of Eq. (56) of \cite{Arroja:2015yvd} for a model that includes also an imperfect fluid. It is worth noting that if the fluid is dust then this equation together with Eq. (\ref{eq6}) imply that the evolution of $\Phi$ and $v$ is scale-invariant even if the background is not the same as in the LCDM model.

If we consider the case of the LCDM background expansion history, then $\tilde\Gamma=0$. In other words, the equation $\tilde\Gamma=0$ can be integrated twice to give
\begin{equation}
3M_{Pl}^2\mathcal{H}^2=a^2\left(\rho_{m0}a^{-3}+\Lambda\right),
\end{equation}
where $\rho_{m0}$ and $\Lambda$ are two integration constants (the matter (baryon plus dark matter) density today and Einstein's cosmological constant).
If we further assume the fluid is a fluid of baryons with $\Pi=\delta p=\bar{p}=0$ and $\bar{\rho}=\bar{\rho}_{b0}a^{-3}$ (note that $\rho_{m0}$ is a constant independent of $\bar{\rho}_{b0}$) then the equation simplifies greatly as
\begin{equation}
\Phi''+3\mathcal{H}\Phi'+\left(\mathcal{H}^2+2\mathcal{H}'\right)\Phi=0.
\end{equation}
The previous equation is the same evolution equation for the potential $\Phi$ that one finds in the LCDM model. So, for identical initial conditions, we will find the same potential today in both models.

\section{The equations of motion in perfect fluid dark energy models\label{app:eomPFDE}}

In this appendix, we briefly summarize the well-known equations of motion in a fluid dark energy model coupled with GR. Perfect fluids have $\Pi_\mathfrak{f}=0$ but here we keep $\Pi$ in the equations for generality.
The conservation of the energy-momentum tensors, assuming the different matter species do not exchange energy, give
\begin{eqnarray}
&&
\delta_\mathfrak{f}'+3\mathcal{H}\left(c_{(\mathfrak{f})s}^2-w_\mathfrak{f}\right)\delta_\mathfrak{f}-3(1+w_\mathfrak{f})\Psi'+\left(1+w_\mathfrak{f}\right)\left[-k^2-9\mathcal{H}^2\left(c_{(\mathfrak{f})s}^2-c_{(\mathfrak{f})a}^2\right)\right]v_\mathfrak{f}=0,
\label{deltap}
\\
&&
v_\mathfrak{f}'+\mathcal{H}\left(1-3c_{(\mathfrak{f})s}^2\right)v_\mathfrak{f}+\frac{c_{(\mathfrak{f})s}^2}{1+w_\mathfrak{f}}\delta_\mathfrak{f}+\Phi-\frac{2k^2}{3\bar{\rho}_\mathfrak{f}\left(1+w_\mathfrak{f}\right)}\Pi_\mathfrak{f}=0.
\label{vp}
\end{eqnarray}
The following two equations close the system,
\begin{equation}
M_{Pl}^2(\Psi-\Phi)=a^2\Pi, \quad \Psi'+\mathcal{H}\Phi+\frac{a^2}{2M_{Pl}^2}\left(\bar{\rho}+\bar{p}\right)v=0,
\label{EEPF}
\end{equation}
where
\begin{equation}
\Pi\equiv\sum_\mathfrak{f} \Pi_\mathfrak{f}, \quad \left(\bar{\rho}+\bar{p}\right)v\equiv\sum_\mathfrak{f} \left(\bar{\rho}_\mathfrak{f}+\bar{p}_\mathfrak{f}\right)v_\mathfrak{f}.
\end{equation}
The first and second equations come from the ij and 0i parts of the Einstein equations, respectively. Note that we do not need to use the other ij equation. The 00 equation, known as the Poisson equation, is a constraint and reads
\begin{equation}
2M_{Pl}^2\partial^2\Psi=a^2\sum_\mathfrak{f}\bar{\rho}_\mathfrak{f}\left(\delta_\mathfrak{f}-3(1+w_\mathfrak{f})v_\mathfrak{f}\right).
\end{equation}

In the background, we have
\begin{equation}
\mathcal{H}^2=a^2H_0^2\left(\Omega_{m0}a^{-3}+\Omega_{r0}a^{-4}+\Omega_{DE0}e^{3(a-1)w_a}a^{-3(1+w_0+w_a)}\right).
\end{equation}
For the CPL parametrization, the adiabatic sound speed is $c_a^2=(w_a a+3w_{DE}+3w_{DE}^2)/(3(1+w_{DE}))$. Also one has $1=\Omega_m+\Omega_r+\Omega_{DE}$. We use the number of efolds, $N=\ln a$, as the time variable ($a_0=1$).

At the linear order, we have the following complete set of equations that we integrate numerically
\begin{eqnarray}
&&\Psi=\Phi,
\\
&&\frac{d\Phi}{dN}+\Phi=-\frac{3}{2}\frac{a^2H_0^2}{\mathcal{H}^2}\left(\Omega_{m0}a^{-3}\tilde{v}_m+\frac{4}{3}\Omega_{r0}a^{-4}\tilde{v}_r+\Omega_{DE0}(1+w_{DE})e^{3(a-1)w_a}a^{-3(1+w_0+w_a)}\tilde{v}_{DE}\right),
\label{eomfluidDE}
\\
&&\frac{d\delta_{DE}}{dN}+3\left(c_s^2-w_{DE}\right)\delta_{DE}+(1+w_{DE})\left[-9(c_s^2-c_a^2)-\frac{k^2}{\mathcal{H}}\right]\tilde{v}_{DE}-3(1+w_{DE})\frac{d\Phi}{dN}=0,
\\
&&\frac{d\tilde{v}_{DE}}{dN}+\left[1-3c_s^2-\frac{\mathcal{H}'}{\mathcal{H}^2}\right]\tilde{v}_{DE}+\Phi+\frac{c_s^2}{1+w_{DE}}\delta_{DE}=0,
\\
&&\frac{d\delta_m}{dN}-3\frac{d\Phi}{dN}-\frac{k^2}{\mathcal{H}^2}\tilde{v}_m=0,
\\
&&\frac{d\tilde{v}_m}{dN}+\left(1-\frac{\mathcal{H}'}{\mathcal{H}^2}\right)\tilde{v}_m+\Phi=0,
\\
&&\frac{d\delta_r}{dN}-4\frac{d\Phi}{dN}-\frac{4}{3}\frac{k^2}{\mathcal{H}^2}\tilde{v}_r=0,
\\
&&\frac{d\tilde{v}_r}{dN}-\frac{\mathcal{H}'}{\mathcal{H}^2}\tilde{v}_r+\Phi+\frac{\delta_r}{4}=0,
\end{eqnarray}
where $k$ is the wavenumber of Fourier space.

\section{The power spectrum of $\Phi$ in clustering perfect fluid dark energy models\label{app:PSPFDE}}

In this appendix, we briefly introduce the transfer function of \cite{Eisenstein:1997ik} and present a fitting function approximation for the power spectrum of the Newtonian potential in the LCDM model. We then turn to the main goal of this appendix and discuss a fitting function approximation for the power spectrum of $\Phi$ in clustering perfect fluid dark energy models (where gravity is assumed to be described by GR) following e.g. \cite{Hu:1998tj,Eisenstein:1997ik,Hu:2001fb,Takada:2006xs}.

The power spectrum of $\Phi$ can be written at a certain redshift, $z$, as \cite{Eisenstein:1997ik} (EH stands for Eisenstein and Hu)
\begin{equation}
P^{EH(BAO)}_\Phi(k,z)=\left(\frac{3}{5}\right)^2\Delta_\mathcal{R}^2(k)T(k)^2\left(\frac{g(z)}{g_\infty}\right)^2,\label{EH}
\end{equation}
where the transfer function (Eq. (16) of \cite{Eisenstein:1997ik}) is
\begin{equation}
T(k)=\frac{\Omega_{b0}}{\Omega_{m0}}T_b(k)+\frac{\Omega_{DM0}}{\Omega_{m0}}T_{DM}(k).\label{transferfc}
\end{equation}
We will not reproduce here the analytical expressions for the transfer functions of baryons and dark matter. See their paper for all the details. In this work we use the transfer function with baryon acoustic oscillations (BAO) and they assume adiabatic perturbations.
These transfer functions are for baryon (neutralized by accompanying electrons) and cold dark matter in a universe also composed of photons and massless neutrinos (and antineutrinos). The existence of a non-zero cosmological constant today (or recently) or spatial curvature is insignificant.
These transfer functions agree with the exact results from Boltzmann codes to better than $5\%$.

While essentially the transfer function describes the evolution of the gravitational potential on sub-horizon scales during the radiation era, to describe the evolution in the recent dark energy dominated era one needs to introduce the growth factor.

The growth factor in the $\Lambda$CDM model is \cite{Lahav:1991wc,Carroll:1991mt}
\begin{eqnarray}
g(z)&=&\frac{5}{2}\Omega_m(z)g_\infty\left[\Omega_m(z)^{4/7}-\Omega_\Lambda(z)+\left(1+\frac{\Omega_m(z)}{2}\right)\left(1+\frac{1}{70}\Omega_\Lambda(z)\right)\right]^{-1},
\\
\Omega_m(z)&=&\frac{\Omega_{m0}(1+z)^3}{\Omega_{m0}(1+z)^3+(1-\Omega_{m0}-\Omega_{r0})},\quad
\Omega_\Lambda(z)=\frac{1-\Omega_{m0}-\Omega_{r0}}{\Omega_{m0}(1+z)^3+(1-\Omega_{m0}-\Omega_{r0})}.
\end{eqnarray}

Let us now turn to the main goal of this appendix, i.e. to  present an approximation for the power spectrum of $\Phi$ in clustering perfect fluid dark energy models.

If one assumes that the dark energy is a perfect fluid then at the perturbation level there is an additional free parameter, the sound speed $c_s$. If $c_s\approx1$ then the dark energy will develop perturbations only on the horizon scale which is beyond current observational efforts in LSS surveys.
However if $c_s$ is smaller than one or it is zero then dark energy perturbations will be non-zero on smaller scales or even on all scales for the case $c_s=0$.

One defines the sound horizon of the dark energy as
\begin{equation}
\lambda_{DE}(a)=\int_0^a da \frac{c_s}{a^2H}.
\end{equation}
For scales larger than the sound horizon, in the so-called clustering regime, dark energy can cluster and
there exists a well-know solution (growing mode) \cite{Hu:1998tj}
\begin{equation}
\Phi_c\propto1-\frac{H}{a}\int \frac{da}{H},\label{phicsol}
\end{equation}
where the proportionality constant is determined by the initial condition.

For scales much smaller than the sound horizon during the dark energy era, the dark energy perturbations are negligible (i.e. the dark energy component is smooth at all times). In this case, setting $v_{DE}=0$, the system of equations reads
\begin{eqnarray}
\Phi=\Psi, \quad \Psi'+\mathcal{H}\Phi+\frac{a^2}{2M_{Pl}^2}\bar{\rho}_m v_m=0,
\quad
v_m'+\mathcal{H}v_m+\Phi=0,
\end{eqnarray}
which can be combined to find
\begin{equation}
\Phi''+3\mathcal{H}\Phi'+\frac{3}{2}\mathcal{H}^2\Omega_{DE}(a)\left(1-w_{DE}(a)\right)\Phi=0,
\end{equation}
where the equation of state $w_{DE}(a)$ is general. We solve this equation numerically (in terms of the number of efolds) and denote its solution by $\Phi_s$, with initial conditions set deep in the matter dominated era as $\Phi_s(a_{md})=3/5,\,\Phi(a_{md})'=0$, where $a_{md}=10^{-3}$ for example.

In the case of $w_a=0$, the power spectrum of the potential perturbation can be written as
\begin{equation}
P_\Phi^{EHT}(k,z)=\Delta_\mathcal{R}^2(k)T_{DE}(k,z)^2\Phi_s(z)^2T(k)^2,\label{EHT}
\end{equation}
where the primordial spectrum, $\Delta_\mathcal{R}^2(k)$, and the transfer function, $T(k)$, were defined in Eqs. (\ref{primordialspectrum}) and (\ref{transferfc}) respectively.
Following \cite{Hu:2001fb}, the interpolation function $T_{DE}(k,z)$ can be written as
\begin{equation}
T_{DE}(k,z)=\frac{1+\bar{q}^2}{\Phi_s/\Phi_c+\bar{q}^2},\label{TDE}
\end{equation}
where
\begin{equation}
\bar{q}\equiv\frac{k}{2\pi}\sqrt{\lambda_{DE}(z)\lambda_{DE}(z_{DE})},
\end{equation}
with the redshift $z_{DE}$ defined as
\begin{equation}
\frac{\bar{\rho}_{DE}(z_{DE})}{\bar{\rho}_m(z_{DE})}=\frac{1}{\pi},\quad 1+z_{DE}=\left(\pi\frac{\Omega_{DE0}}{\Omega_{m0}}\right)^{-\frac{1}{3w_0}}.
\end{equation}

It is worth noting that in the limit, $c_s=0$, the power spectrum takes a simple form
\begin{equation}
P_\Phi^{EHT\,c_s=0}(k,z)=\Delta_\mathcal{R}^2(k)\Phi_c(z)^2T(k)^2,
\end{equation}
where one has an analytical solution for $\Phi_c$ as given in equation (\ref{phicsol}) (taking the integration constant equal to one when inserting it in the previous expression for the power spectrum). Ref. \cite{Hu:2001fb}, showed that the fitting formula (\ref{TDE}) approximates well (relative error of about $10\%$) the exact numerical results of a multi-fluid Boltzmann code with dark energy clustering.

\bibliography{draftPublished.bbl}

\begin{thebibliography}{10}

\bibitem{Dodelson:2003ft}
S.~Dodelson,
\newblock {\em {Modern Cosmology}} (Academic Press, Amsterdam, 2003).

\bibitem{Weinberg:2008zzc}
S.~Weinberg,
\newblock {\em {Cosmology}} (Oxford Univ. Press, Oxford, UK, 2008).

\bibitem{Adam:2015rua}
Planck, R.~Adam {\em et~al.},
\newblock Astron. Astrophys. {\bf 594}, A1 (2016), 1502.01582.

\bibitem{Ade:2015xua}
Planck, P.~A.~R. Ade {\em et~al.},
\newblock Astron. Astrophys. {\bf 594}, A13 (2016), 1502.01589.

\bibitem{Riess:1998cb}
Supernova Search Team, A.~G. Riess {\em et~al.},
\newblock Astron. J. {\bf 116}, 1009 (1998), astro-ph/9805201.

\bibitem{Perlmutter:1998np}
Supernova Cosmology Project, S.~Perlmutter {\em et~al.},
\newblock Astrophys. J. {\bf 517}, 565 (1999), astro-ph/9812133.

\bibitem{Weinberg:1988cp}
S.~Weinberg,
\newblock Rev. Mod. Phys. {\bf 61}, 1 (1989).

\bibitem{Bull:2015stt}
P.~Bull {\em et~al.},
\newblock Phys. Dark Univ. {\bf 12}, 56 (2016), 1512.05356.

\bibitem{Tsujikawa:2013fta}
S.~Tsujikawa,
\newblock Class. Quant. Grav. {\bf 30}, 214003 (2013), 1304.1961.

\bibitem{Frieman:2008sn}
J.~Frieman, M.~Turner, and D.~Huterer,
\newblock Ann. Rev. Astron. Astrophys. {\bf 46}, 385 (2008), 0803.0982.

\bibitem{Clifton:2011jh}
T.~Clifton, P.~G. Ferreira, A.~Padilla, and C.~Skordis,
\newblock Phys. Rept. {\bf 513}, 1 (2012), 1106.2476.

\bibitem{Joyce:2016vqv}
A.~Joyce, L.~Lombriser, and F.~Schmidt,
\newblock Ann. Rev. Nucl. Part. Sci. {\bf 66}, 95 (2016), 1601.06133.

\bibitem{Hui:2016ltb}
L.~Hui, J.~P. Ostriker, S.~Tremaine, and E.~Witten,
\newblock Phys. Rev. {\bf D95}, 043541 (2017), 1610.08297.

\bibitem{Chamseddine:2013kea}
A.~H. Chamseddine and V.~Mukhanov,
\newblock JHEP {\bf 1311}, 135 (2013), 1308.5410.

\bibitem{Chamseddine:2014vna}
A.~H. Chamseddine, V.~Mukhanov, and A.~Vikman,
\newblock JCAP {\bf 1406}, 017 (2014), 1403.3961.

\bibitem{Lim:2010yk}
E.~A. Lim, I.~Sawicki, and A.~Vikman,
\newblock JCAP {\bf 1005}, 012 (2010), 1003.5751.

\bibitem{Sebastiani:2016ras}
L.~Sebastiani, S.~Vagnozzi, and R.~Myrzakulov,
\newblock Adv. High Energy Phys. {\bf 2017}, 3156915 (2017), 1612.08661.

\bibitem{Arroja:2015wpa}
F.~Arroja, N.~Bartolo, P.~Karmakar, and S.~Matarrese,
\newblock JCAP {\bf 1509}, 051 (2015), 1506.08575.

\bibitem{Horndeski:1974wa}
G.~W. Horndeski,
\newblock Int.J.Theor.Phys. {\bf 10}, 363 (1974).

\bibitem{Langlois:2017mdk}
D.~Langlois,
\newblock {Degenerate Higher-Order Scalar-Tensor (DHOST) theories},
\newblock in {\em {52nd Rencontres de Moriond on Gravitation (Moriond
  Gravitation 2017) La Thuile, Italy, March 25-April 1, 2017}}, 2017,
  1707.03625.

\bibitem{Arroja:2015yvd}
F.~Arroja, N.~Bartolo, P.~Karmakar, and S.~Matarrese,
\newblock JCAP {\bf 1604}, 042 (2016), 1512.09374.

\bibitem{Ramazanov:2016xhp}
S.~Ramazanov, F.~Arroja, M.~Celoria, S.~Matarrese, and L.~Pilo,
\newblock JHEP {\bf 06}, 020 (2016), 1601.05405.

\bibitem{Horava:2009uw}
P.~Horava,
\newblock Phys. Rev. {\bf D79}, 084008 (2009), 0901.3775.

\bibitem{Capela:2014xta}
F.~Capela and S.~Ramazanov,
\newblock JCAP {\bf 1504}, 051 (2015), 1412.2051.

\bibitem{Bertacca:2010ct}
D.~Bertacca, N.~Bartolo, and S.~Matarrese,
\newblock Adv. Astron. {\bf 2010}, 904379 (2010), 1008.0614.

\bibitem{Ellis:2012rn}
PFS Team, R.~Ellis {\em et~al.},
\newblock Publ. Astron. Soc. Jap. {\bf 66}, R1 (2014), 1206.0737.

\bibitem{Aghamousa:2016zmz}
DESI, A.~Aghamousa {\em et~al.},
\newblock (2016), 1611.00036.

\bibitem{Amendola:2016saw}
L.~Amendola {\em et~al.},
\newblock (2016), 1606.00180.

\bibitem{Matsumoto:2016rsa}
J.~Matsumoto,
\newblock (2016), 1610.07847.

\bibitem{Lewis:1999bs}
A.~Lewis, A.~Challinor, and A.~Lasenby,
\newblock Astrophys. J. {\bf 538}, 473 (2000), astro-ph/9911177.

\bibitem{Deffayet:2011gz}
C.~Deffayet, X.~Gao, D.~A. Steer, and G.~Zahariade,
\newblock Phys. Rev. {\bf D84}, 064039 (2011), 1103.3260.

\bibitem{Kobayashi:2011nu}
T.~Kobayashi, M.~Yamaguchi, and J.~Yokoyama,
\newblock Prog. Theor. Phys. {\bf 126}, 511 (2011), 1105.5723.

\bibitem{Deruelle:2014zza}
N.~Deruelle and J.~Rua,
\newblock JCAP {\bf 1409}, 002 (2014), 1407.0825.

\bibitem{Chevallier:2000qy}
M.~Chevallier and D.~Polarski,
\newblock Int. J. Mod. Phys. {\bf D10}, 213 (2001), gr-qc/0009008.

\bibitem{Linder:2002et}
E.~V. Linder,
\newblock Phys. Rev. Lett. {\bf 90}, 091301 (2003), astro-ph/0208512.

\bibitem{Kodama:1985bj}
H.~Kodama and M.~Sasaki,
\newblock Prog. Theor. Phys. Suppl. {\bf 78}, 1 (1984).

\bibitem{Bean:2003fb}
R.~Bean and O.~Dore,
\newblock Phys. Rev. {\bf D69}, 083503 (2004), astro-ph/0307100.

\bibitem{Deffayet:2010qz}
C.~Deffayet, O.~Pujolas, I.~Sawicki, and A.~Vikman,
\newblock JCAP {\bf 1010}, 026 (2010), 1008.0048.

\bibitem{Eisenstein:1997ik}
D.~J. Eisenstein and W.~Hu,
\newblock Astrophys. J. {\bf 496}, 605 (1998), astro-ph/9709112.

\bibitem{Hu:2001fb}
W.~Hu,
\newblock Phys. Rev. {\bf D65}, 023003 (2002), astro-ph/0108090.

\bibitem{Takada:2006xs}
M.~Takada,
\newblock Phys. Rev. {\bf D74}, 043505 (2006), astro-ph/0606533.

\bibitem{Okumura:2011pb}
T.~Okumura, U.~Seljak, P.~McDonald, and V.~Desjacques,
\newblock JCAP {\bf 1202}, 010 (2012), 1109.1609.

\bibitem{Okumura:2012xh}
T.~Okumura, U.~Seljak, and V.~Desjacques,
\newblock JCAP {\bf 1211}, 014 (2012), 1206.4070.

\bibitem{Desjacques:2008vf}
V.~Desjacques, U.~Seljak, and I.~Iliev,
\newblock Mon. Not. Roy. Astron. Soc. {\bf 396}, 85 (2009), 0811.2748.

\bibitem{Challinor:2011bk}
A.~Challinor and A.~Lewis,
\newblock Phys. Rev. {\bf D84}, 043516 (2011), 1105.5292.

\bibitem{Jeong:2011as}
D.~Jeong, F.~Schmidt, and C.~M. Hirata,
\newblock Phys. Rev. {\bf D85}, 023504 (2012), 1107.5427.

\bibitem{Ade:2015rim}
Planck, P.~A.~R. Ade {\em et~al.},
\newblock Astron. Astrophys. {\bf 594}, A14 (2016), 1502.01590.

\bibitem{Takahashi:2017pje}
K.~Takahashi and T.~Kobayashi,
\newblock JCAP {\bf 1711}, 038 (2017), 1708.02951.

\bibitem{Langlois:2018jdg}
D.~Langlois, M.~Mancarella, K.~Noui, and F.~Vernizzi,
\newblock (2018), 1802.03394.

\bibitem{Hu:1998tj}
W.~Hu and D.~J. Eisenstein,
\newblock Phys. Rev. {\bf D59}, 083509 (1999), astro-ph/9809368.

\bibitem{Lahav:1991wc}
O.~Lahav, P.~B. Lilje, J.~R. Primack, and M.~J. Rees,
\newblock Mon. Not. Roy. Astron. Soc. {\bf 251}, 128 (1991).

\bibitem{Carroll:1991mt}
S.~M. Carroll, W.~H. Press, and E.~L. Turner,
\newblock Ann. Rev. Astron. Astrophys. {\bf 30}, 499 (1992).

\end{thebibliography}

\end{document}